\documentclass[twoside,twocolumn]{article}

\usepackage[sc]{mathpazo} 
\usepackage[T1]{fontenc} 
\usepackage{microtype} 

\usepackage[english]{babel} 

\usepackage{siunitx} 
\usepackage[super]{nth} 

\usepackage{graphicx}

\usepackage[a4paper,top=2cm,bottom=2cm,left=2cm,right=2cm,marginparwidth=2cm]{geometry}

\usepackage[small,labelfont=bf,up]{caption} 
\usepackage{booktabs} 

\usepackage[numbers,compress]{natbib} 
\usepackage{lettrine} 

\usepackage{enumitem} 
\setlist[itemize]{noitemsep} 

\usepackage{abstract} 

\usepackage{titlesec} 
\titleformat{\section}[block]{\large\centering}{\thesection.}{1em}{} 
\titleformat{\subsection}[block]{\large}{\thesubsection.}{1em}{} 

\usepackage{fancyhdr} 
\usepackage{titling} 

\usepackage[colorinlistoftodos]{todonotes}
\usepackage[colorlinks=true, allcolors=blue]{hyperref}
\DeclareUnicodeCharacter{2212}{-} 
\DeclareUnicodeCharacter{0301}{ }

\pretitle{\begin{center}\Huge} 
\posttitle{\end{center}} 
\title{Pump-probe cathodoluminescence microscopy} 
\usepackage{authblk}
\author[1]{Magdalena Solà-Garcia}
\author[1]{Kelly W. Mauser}
\author[1]{Nika van Nielen}
\author[2]{Toon Coenen}
\author[3]{Sophie Meuret}
\author[1]{Albert Polman}
\affil[1]{Center for Nanophotonics, AMOLF, 1098XG Amsterdam, The Netherlands}
\affil[2]{Delmic BV, 2628 EB Delft, The Netherlands}
\affil[3]{CEMES-CNRS, 31055 Toulouse, France}

\date{} 
\renewcommand{\maketitlehookd}{%
\begin{abstract}
\noindent
	\normalsize{
	In this paper we describe the technical implementation of pump-probe cathodoluminescence microscopy (PP-CL), a novel technique for studying ultrafast dynamics in materials using combined excitation of electron and laser pulses. The PP-CL setup is based on a commercial scanning electron microscope combined with a femtosecond laser. Here we discuss the design, technical details, and characterization of the PP-CL setup. The luminescence emitted by the sample (either photoluminescence, cathodoluminescence or a combination of both) can be analyzed in terms of spectrum, temporal statistics, and frequency-modulated intensity (lock-in detection). The laser spot on the sample is characterized by analyzing the changes in the secondary electron signal on a GaN substrate, from which we extract a laser spot size in the \SI{}{\micro\meter} range. While this method ensures spatial overlap of laser and electron pulses on the sample, temporal alignment is achieved using time-correlated measurements of photoluminescence and cathodoluminescence. We also discuss some considerations to take into account when designing a PP-CL experiment, regarding temporal and spatial resolution. This paper provides a detailed reference for the development of any PP-CL microscope.}
\end{abstract}
}

\begin{document}

\maketitle

\section{Introduction}
Cathodoluminescence (CL) microscopy is a powerful technique for resolving optical properties down to the nanometer scale, given the use of electrons as the excitation source. The spatial resolution of CL is limited by the electron beam size ($\sim \SI{}{\nano\meter}$), the interaction volume of the electron beam and the diffusion of charge carriers inside the material. Recently, the emergence of ultrafast electron microscopy has enabled time-resolved CL (TR-CL) studies, in which the emission and excitation dynamics of materials are investigated at the nanometer scale \citep{Merano2005, Sonderegger2006b,Moerland2016a}. The time resolution of TR-CL is typically limited by the detection system ($\sim$ tens of ps) \citep{Meuret2019}. Instead, by combining optical and electron excitation in a pump-probe configuration we can achieve temporal resolutions down to the hundreds of fs regime, limited by the electron and optical pulse durations, while taking advantage of the high spatial resolution given by the electron beam \citep{Zewail2010}.

Pump-probe measurements, usually based on double-laser beam excitation, are routinely used across different research fields, including biochemistry, materials science and molecular physics \citep{Fushitani2008,Cabanillas-Gonzalez2011,Grumstrup2015}. In these experiments, a sample is excited with a pump beam (typically an intense laser beam) and the resulting state is probed with a second beam (for example, another laser pulse), thus enabling the study of ultrafast processes. Even though fully optical pump-probe configurations are the most common ones, many different pump-probe schemes have been proposed, with changes either in one or both of the pump/probe beams \citep{Nakajima2008,Picon2016} or the detection scheme \citep{Murawski2015,Jahng2015}. In electron microscopy, pump-probe measurements using a laser-pump electron-probe configuration have been demonstrated \citep{Lobastov2005,Barwick2008,Mohammed2011}. A pump-probe based technique that is gaining attention is photon-induced near-field electron microscopy (PINEM), typically performed in a (scanning) transmission electron microscope, (S)TEM, which is based on the study of the electron energy loss and gain after interaction with an optically-excited nanostructure \citep{Barwick2009,GarciaDeAbajo2010a,Feist2015}. Other pump-probe-type works in a TEM include the study of the formation of chemical bonds \citep{Carbone2009a}, magnetization dynamics \citep{Schliep2017,RubianoDaSilva2018} and optically-excited phonon modes \citep{Cremons2016,Valley2016}. In scanning electron microscopes (SEM), previous studies have investigated the recombination dynamics in semiconductors of optically-induced carriers by analyzing the variations in the secondary electron signal \citep{Mohammed2011,Liao2017a, Garming2020}. In all these cases the sample is excited by the laser pump, while the electrons act as a probe. The probe signals are thus transmitted electrons (used for real and Fourier-space imaging or EELS, among others) or secondary electrons, from which a real-space image is formed.

In this paper, we discuss the implementation and characterization of the first pump-probe cathodoluminescence (PP-CL) setup. Similar to previous work, our setup consists of a two-beam system, with both pulsed electron and laser beams. In contrast to earlier work, the emission and excitation dynamics are investigated through the analysis of the emitted luminescence, either CL or photoluminescence (PL). Hence, our setup allows us to use the electron beam either as a probe, after pumping with the laser, or as a pump, which is a novelty compared to other pump-probe experiments. This last operating mode was recently used to show electron-induced charge state conversion in nitrogen-vacancy centers in diamond \citep{Sola-Garcia2019}.

\section{Overview of the PP-CL setup}\label{sec:Setup-chapter_Overview}
Our pump-probe experiments rely on the use of an electron and laser beam, in which one (pump) brings the sample out of equilibrium and the other one (probe) records the new state of the material. Tuning the delay between pump and probe gives access to the dynamics of the induced effect. The process of pumping and probing is repeated over many cycles ($>10^6$ s$^{-1}$) in order to accumulate enough signal (stroboscopic mode) \citep{Arbouet2018}. Hence, this method can only be used to study reversible phenomena, given that the sample has to go back to a steady (unexcited) state before each cycle of pump-probe. 

\begin{figure*}
	\centering
	\includegraphics[width=\textwidth]{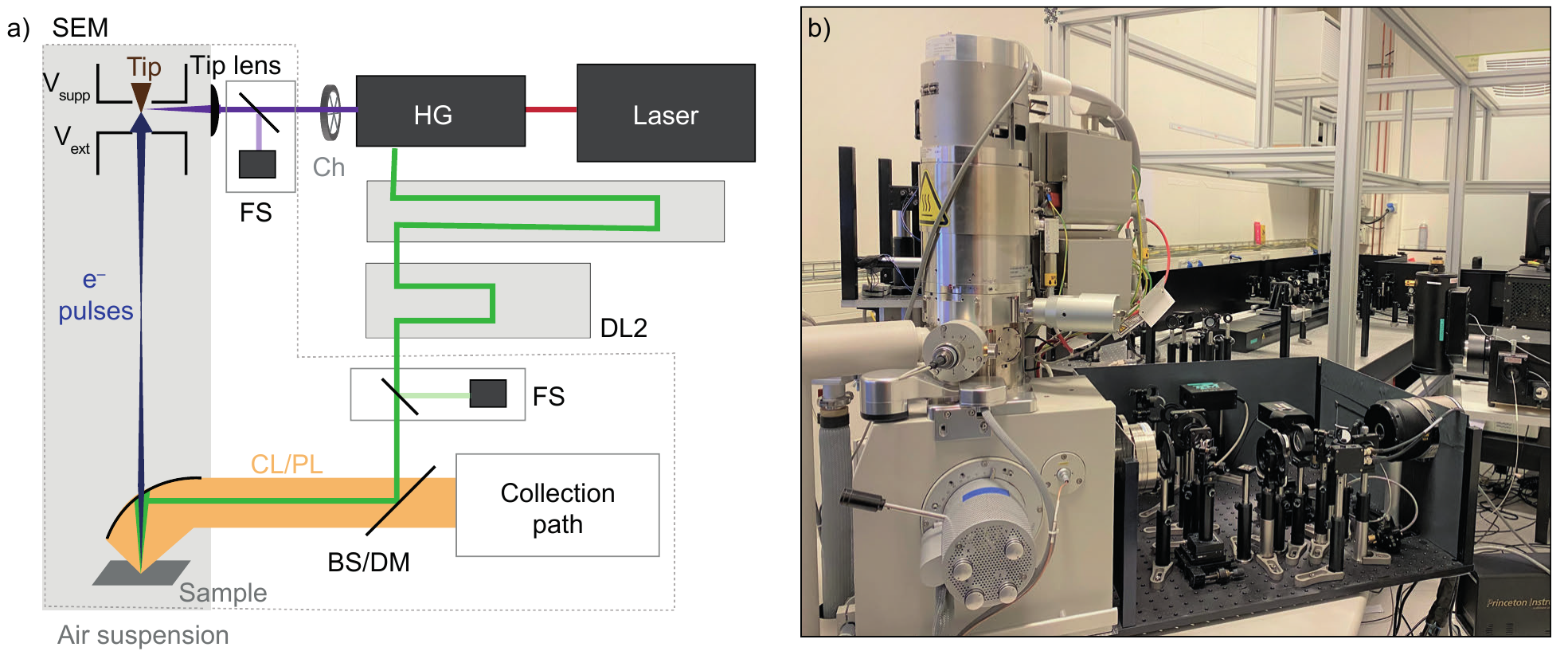}
	\caption{Pump-probe cathodoluminescence setup. Schematic (a) and photograph (b) of our PP-CL setup. The setup consists of the electron (purple), light-injection (green) and collection (orange) paths. The fundamental output ($\lambda=\SI{1035}{\nano\meter}$) of a femtosecond laser is converted to the \nth{2}, \nth{3} and \nth{4} harmonics ($\lambda=517,345$ and \SI{258}{\nano\meter}, respectively) in the harmonic generator (HG) setup. The \nth{4} harmonic is focused on the electron cathode of the SEM to generate electron pulses. The \nth{2} or \nth{3} harmonics are directed towards the SEM chamber using either a beam splitter or dichroic mirror (BS, DM) to optically excite the sample. The delay between electron and light pulses is controlled using two delay lines (DL1 and DL2). The alignment between the air-suspended parts and the optical table is maintained using a feedback system (FS).}
	\label{fig:Setup-chapter_Setup}
\end{figure*}

\begin{figure}
	\centering
	\includegraphics[width=0.45\textwidth]{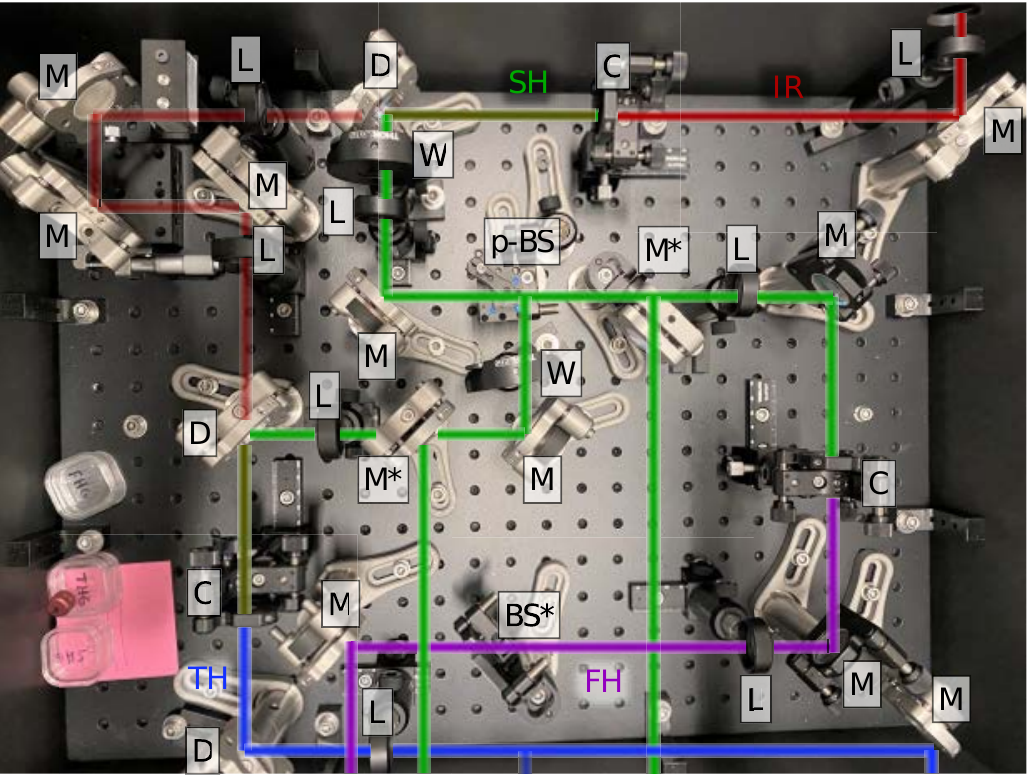}
	\caption{Photograph and beam path of the harmonic generator setup (Clark MXR). The fundamental laser beam (IR, $\lambda=\SI{1035}{\nano\meter}$) is directed towards a set of BBO crystals to generate the \nth{2} (SH, $\lambda=\SI{517}{\nano\meter}$), \nth{3} (TH, $\lambda=\SI{345}{\nano\meter}$) and \nth{4} (FH, $\lambda=\SI{258}{\nano\meter}$) harmonics. This setup allows us to use different harmonics simultaneously. Abbreviations: mirror (M), lens (L), dichroic mirror (D), beam-splitter (BS), polarizing beam-splitter (p-BS) and BBO crystal (C). The asterisk denotes the components that can be flipped in and out of the beam path.}
	\label{fig:Setup-chapter_HG}
\end{figure}

Our pump-probe CL setup integrates an SEM with an optical setup containing a femtosecond laser (Fig. \ref{fig:Setup-chapter_Setup}). We use a Thermo Fisher Quanta 250 FEG SEM ($0.5-\SI{30}{\kilo\electronvolt}$), which has been modified to give optical access to the electron gun through a UV-transparent window. The femtosecond laser (Clark MXR) consists of a diode-pumped Yb-doped fiber oscillator/amplifier system, providing $\sim \SI{250}{\femto\second}$ pulses at an output wavelength of $\sim\SI{1035}{\nano\meter}$ and tunable repetition rate (\SI{200}{\kilo\hertz}-\SI{25.19}{\mega\hertz}). The laser beam is directed towards an optical setup (Clark MXR) containing a set of beta barium borate (BBO) crystals to obtain the \nth{2}, \nth{3} and \nth{4} harmonics of the fundamental beam ($\lambda=517$, $345$ and \SI{258}{\nano\meter}, respectively). Figure \ref{fig:Setup-chapter_HG} shows an image of the harmonic generator (HG) setup with the corresponding beam paths for the different harmonics. The HG setup is built such that we can use different combinations of two harmonics simultaneously, thus offering a large flexibility in a pump-probe experiment. For experiments requiring a larger range of excitation wavelengths, the HG setup could be replaced by an optical parametric amplifier (OPA).

The PP setup can be divided into three main parts: generation of electron pulses, optical excitation of the sample, and collection path. The generation of electron pulses includes the focusing of one of the harmonics of the laser beam (typically the \nth{4} harmonic) on the electron cathode to induce photoemission of electrons, and the subsequent path of electrons inside the microscope column. The optical excitation of the sample refers to the set of optical components designed to direct the laser beam towards the SEM chamber and focus it on the sample. Finally, the collection path denotes the set of optics and detection systems that we use to analyze the luminescence from the sample (both PL and CL). 

The SEM, the collection path and part of the path to generate electron pulses are mounted on an air suspension system (see Fig. \ref{fig:Setup-chapter_Setup}a), meaning that they move freely to compensate for vibrations in the room. In contrast, the rest of optical components, including the laser, is mounted on a non-floating optical table. We use two identical optical feedback systems (TEM-Messtechnik $\mu$-Aligna) for the electron and light-injection paths, respectively, to actively track the movement of the floating section and move the laser beam accordingly, such that the alignment between both sections is maintained. The feedback system is composed of a position-sensitive detector (PSD) and two motorized mirror mounts. The PSD and one of the mirrors are mounted on the floating section, while the additional mirror is placed along the non-floating part (either the electron generation or the sample excitation paths). All components are connected through a controller. We use a beam sampler to send a small fraction of the laser beam to the PSD, which tracks the motion of the SEM with respect to the laser beam, and the mechanical actuators are moved accordingly to compensate for it.

We use an off-axis parabolic mirror (\SI{0.5}{\milli\meter} focal distance, parabola coefficient a=0.1 and $1.48\pi$ sr solid angle collection) positioned above the sample both to focus the laser beam on the sample and to collect the luminescence (PL or CL) \citep{Coenen2011}. The mirror has a \SI{600}{\micro\meter} diameter hole drilled above the focal point through which electrons pass. A motorized stage (Delmic Redux) controls the position of the mirror in the direction parallel to the sample plane. We bring the sample to the focal point of the mirror by moving the sample stage up to a working distance for the SEM of around \SI{14}{\milli\meter} (see section \ref{sec:Setup-chapter_CCD}). In the next sections we will provide a detailed description of the coupling and alignment of the laser beam on the SEM gun chamber (generation of electron pulses) and chamber (optical excitation of the sample), as well as the detection and analysis of the emitted CL and PL (collection path). 

\section{Generation of electron pulses}
The SEM is equipped with a Schottky field-emission gun (FEG), with geometry shown in Figure \ref{fig:Setup-chapter_Setup}a (left). It consists of a ZrO/W electron cathode surrounded by a negatively-biased voltage usually referred to as suppressor ($V_\textrm{supp}\sim\SI{-500}{\volt}$) to ensure emission of electrons only from the apex of the tip. In normal (continuous) emission, the tip is heated up to $\sim\SI{1800}{\kelvin}$ and a positive extractor voltage ($V_\textrm{ext}\sim\SI{4550}{\volt}$) is applied to prompt emission of electrons by means of the Schottky effect. In contrast, in the PP-CL microscope we use a pulsed electron beam generated through the process of photoemission. Photoemission of electrons using a FEG source has been already reviewed in previous work \citep{Feist2017a,Meuret2019,Sola-Garcia2021a}. In brief, pulsed emission of electrons is induced by focusing a fs laser beam with photon energy above the ZrO/W work function ($\sim\SI{2.9}{\electronvolt}$ \citep{Bronsgeest2009,Feist2017a}) on the electron cathode. In this case, the temperature of the cathode is lowered to $\sim\SI{990}{\kelvin}$ to completely suppress continuous emission of electrons.


We typically use the \nth{4} harmonic ($\lambda=\SI{258}{\nano\meter}$) of the fundamental laser beam to induce photoemission of electrons. The \nth{3} harmonic of the laser beam can also be used, given that the photon energy is above the work function of the ZrO/W cathode. The beam is initally directed towards a beam expander composed of two lenses  (focal length 10 and \SI{50}{\centi\meter}, respectively) to increase the beam size of the \nth{4} harmonic beam by a factor 5. The beam is further guided up to the height of the electron cathode through a periscope. The top part of the periscope contains a dichroic mirror which reflects the \nth{4} harmonic, while transmitting at longer wavelengths. Finally, the beam is focused on the electron cathode through a lens (tip lens, $f=\SI{12.5}{\milli\meter}$) down to a $\sim \SI{10}{\micro\meter}$ spot. The position of the lens in the direction of the optical axis ($z$) is adjusted using a manually controlled linear stage, while the position in the transverse direction ($x,y$) is controlled by means of two motorized stages (PI M-227). 

A critical step in performing time-resolved experiment with an electron microscope is the proper alignment of the laser beam on the electron cathode. The next sections show the typical procedure performed during the installation of the setup and prior to starting an experiment.

\subsection{Initial alignment of the laser}
The first alignment of the laser on the tip can be performed with the help of a CMOS camera and a set of irises. In normal operating conditions, the electron cathode is set to \SI{1800}{\K} and its blackbody radiation can be observed by eye. This radiation is collected by the lens in front of the SEM window ($f=\SI{12.5}{\milli\meter}$) and imaged on a CMOS camera (Thorlabs DCC1645C). In our case, we placed the CMOS camera behind the dichroic mirror from the periscope, and an achromatic lens ($f=\SI{3}{\centi\meter}$) was used to focus the light onto the camera. A photograph of the visible blackbody radiation of the tip is shown in Fig. \ref{fig:Photoemission-chapter_Tip}a. This configuration allows us to roughly align the lens focus on the electron cathode. The focus is later optimized by maximizing the current emitted by the cathode, as explained below. In order to align the laser on the tip (in the transverse direction, $xy$), we placed a set of two irises through which the emitted light goes. Then the laser is aligned such that it goes through the center of the irises, thus giving a rough alignment of the laser on the tip. The exact position of the laser beam with respect to the tip is optimized by scanning the lens in front of the tip with two motorized stages, as will be explained in the next section. We should note that this initial alignment procedure is only needed during the installation of the setup or after major changes in the laser beam path. 

\subsection{Scanning of the lens}\label{sec:Photoemission-chapter_Tip}
Once the photoemission setup is fully installed, the steps presented in the previous section can be omitted. Hence, we only need to fine-tune the position of the laser on the electron cathode within a $\sim 100\times \SI{100}{\micro\meter\squared}$ scanning window. It is usually desirable to start the alignment of the laser beam on the tip with a hot tip, that is, using the standard settings of the electron column. When lowering the temperature of the tip, it thermally contracts by up to a few tens of \SI{}{\micro\meter} \citep{Bronsgeest2009}, which results in a change of the alignment of the electron beam path inside the column. Hence, starting the alignment of the laser on the cathode with a hot tip ensures an optimum collection of the photoemitted electrons. It is also helpful to optimize the parameters inside the electron column (gun tilt/shift, condenser lens voltage and aperture) for maximum collection of the electrons in continuous mode. We collect the electron current on the sample by focusing the electron beam on a Faraday cup, placed on the sample stage, which is connected to a picoammeter (Keithley 6485) or to a lock-in amplifier (Zurich Instruments MFLI, \SI{500}{\kilo\hertz}/\SI{5}{\mega\hertz}) through a current amplifier (Femto DLPCA-200). When the tip is hot (that is, working in continuous mode) most of the collected electron current comes from the continuous emission, even if the laser beam is already focused on the tip. However, modulating the \nth{4} harmonic laser beam using an optical chopper (Hz - kHz) (Thorlabs MC2000B-EC), which is also connected to a lock-in amplifier, allows us to discern between continuous (Schottky emission) and pulsed (photoemission) electron currents. Additionally, it can be helpful to further tune the parameters inside the electron column (gun tilt/shift, condenser lens voltage and aperture) in order to maximize the collection of the photoemitted electrons.

\begin{figure}
	\centering
	\includegraphics[width=0.45\textwidth]{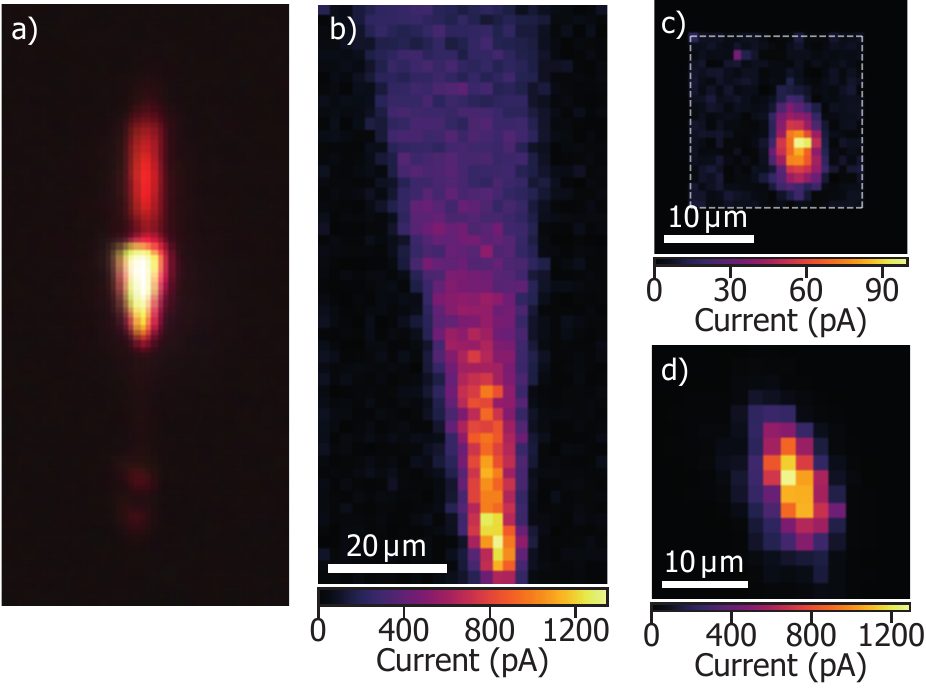}
	\caption{Experimental procedure to focus the laser on the electron cathode. (a) Photograph of blackbody radiation of the tip under normal operating conditions ($T\simeq\SI{1800}{\kelvin}$). (b) Map of the pulsed current collected on a Faraday cup while scanning the laser focus around the hot tip in the $xy$ plane ($T=\SI{1800}{\kelvin}$, $V_{ext}=\SI{4550}{\volt}$). The current produced by photoemission is distinguished from the continuous one using a lock-in amplifier. (c) Similar map as in (b) but with a colder tip ($T\simeq\SI{1200}{\kelvin}$). (d) Similar map as in (c) but obtained at $V_{ext}=\SI{650}{\volt}$. Here the continuous emission is fully suppressed and the pulsed current is measured directly with a picoammeter.}
	\label{fig:Photoemission-chapter_Tip}
\end{figure}

Figure \ref{fig:Photoemission-chapter_Tip}b shows a measurement obtained when scanning the tip lens around the electron cathode (in the $xy$ plane) and collecting the electron signal from the lock-in amplifier. In this case the Schottky FEG was operating in normal conditions (\SI{1800}{\K}, $V_{\textrm{ext}}=\SI{4550}{\volt}$), the electron beam acceleration was set to \SI{10}{\kilo\volt} and the continuous (background) electron current was $\sim\SI{200}{\nano\ampere}$ (using a \SI{1}{\milli\meter} aperture). We used a laser power ($\lambda_{\textrm{FH}}=\SI{258}{\nano\meter}$) of \SI{1.6}{\milli\watt} at \SI{5.04}{\mega\hertz} (\SI{0.3}{\nano\joule}/pulse). The gain of the current amplifier was set to $10^6$ \SI{}{\ampere\per\volt} and the chopping frequency was \SI{287}{\hertz}. We observe that photoemitted current is collected even when the laser is focused more than \SI{60}{\micro\meter} away from the apex of the tip. The configuration of suppressor and extractor voltage ensures that only electrons emitted from the apex of the tip can be released and go through the extractor aperture, and it is thus unlikely that electrons far from the apex (shank emission) can be efficiently released. Instead, the current observed when focusing far from the apex of tip is probably due to emission of electrons from the apex that are excited by the tail of the laser beam profile (assumed to be Gaussian). 

Once the laser is aligned on the tip, we lower the temperature of the tip to suppress continuous emission. This is done by decreasing the filament current from \SI{2.35}{\ampere} to \SI{1.7}{\ampere}, resulting in a final temperature of $\sim \SI{1200}{\kelvin}$. Figure \ref{fig:Photoemission-chapter_Tip}c shows a scan of the tip lens obtained under the same conditions as in \ref{fig:Photoemission-chapter_Tip}b, but at this lower temperature. Given the thermal contraction of the tip at this low temperature, here we realigned the electron column and adjusted the condenser voltage ($C_1$) for optimal electron collection. Here, $C_1$ was increased from \SI{1125}{\volt} in normal conditions to \SI{1190}{\volt}. We observe that the overall emission decreased by $\sim\SI{92}{\percent}$, down to the tens of pA range. Moreover, now emission can be observed only from a relatively small area, corresponding to the apex of the tip, due to the lower photoemission efficiency. Regardless of this reduction in the temperature of the tip, complete suppression of continuous emission is only achieved after letting the tip cool down for at least \SI{1}{\hour}, which is not practical for experiments.

A way to instantaneously remove the remaining continuous emission is by lowering the extractor voltage, such that the effective work function of the ZrO/W tip is increased. We have observed that lowering the extractor voltage to $\sim \SI{3000}{\volt}$ results in the full suppression of continuous emission. Given that the photon energy of the laser beam is much larger than the work function, photoemission of electron pulses is barely affected. Further reducing the extractor voltage to \SI{650}{\volt} results in a change in the angular spread of the photoemitted electrons, leading to a larger collection of electrons \citep{Bronsgeest2009,Sola-Garcia2021a}. In normal operating conditions, the extractor-suppressor configuration acts such that a large fraction of the emitted electrons are blocked by the extractor aperture. Instead, when the magnitude of the extractor voltage is similar to that of the suppressor, most of the emitted electrons can go though the aperture, thus increasing the collection efficiency of the photoemitted electrons \citep{Bronsgeest2009}. We should note that this larger collection efficiency comes at the expense of spatial resolution, given that we collect electrons emitted from a larger area on the source \citep{Meuret2019,Sola-Garcia2021a}. Figure \ref{fig:Photoemission-chapter_Tip}d shows a scan of the tip lens obtained at $V_{\textrm{ext}}=\SI{650}{\volt}$. Given that there is not continuous emission anymore, the laser beam is not chopped and the electron current collected by the Faraday cup is sent directly to a picoammeter. Here again, the condenser lens was readjusted ($V_\textrm{C1}=\SI{650}{\volt}$) together with gun tilt and shift to maximize the collection of electrons. All other parameters were kept the same as in Fig. \ref{fig:Photoemission-chapter_Tip}c. We observe that in this case the maximum electron current is \SI{1.29}{\nano\ampere}, corresponding to an average of $\sim 1600$ electrons per pulse. We should note that this corresponds to one of the lowest values of the extractor voltage at which photoemission is still possible. Lowering $V_{\textrm{ext}}$ below the magnitude of the suppressor voltage (\SI{500}{\volt}) would result in the total suppression of emission of electrons from the tip.

\section{Optical excitation of the sample} \label{sec:Setup-chapter_PLpath}
In our pump-probe setup, we excite the sample with both electron and laser pulses. Hence, we need to guide one of the harmonics of the fs laser towards the sample. Here we use either the \nth{2} or \nth{3} harmonic beams ($\lambda=517$ and \SI{345}{\nano\meter}, respectively) to optically excite the sample. The \nth{4} harmonic ($\lambda=\SI{258}{\nano\meter}$) and fundamental laser beam ($\lambda=\SI{1035}{\nano\meter}$) could also be used to excite the sample by choosing appropriate optics. Figure \ref{fig:Setup-chapter_Setup} shows a schematic of the complete setup, in which the green line represents the laser path for sample excitation. The path is also designed to control the arrival time of the light pulses with respect to that of the electron pulses, which is essential in a pump-probe experiment. We use two free-space optical delay lines (DL1 and DL2 in the figure), consisting of a set of two mirrors (in the case of DL1) or a hollow retroreflector (Newport UBR2.5-5UV, for DL2) mounted on a mechanical stage. Moving the delay stage by \SI{15}{\centi\meter} corresponds to a time delay of \SI{1}{\nano\second}, given the double path of the light along the stage.

The first delay line (DL1) is manually controlled and is used to coarsely adjust the zero-delay between electron and light pulses. We only align this stage at the start of an experiment, and it is kept fixed during a measurement. The operating electron energy of the SEM ($0.5-\SI{30}{\kilo\electronvolt}$) determines the electron arrival time on the sample. Hence, we adjust DL1 in each experiment such that the arrival time of the laser on the sample matches the one from the electrons. For example, a \SI{30}{\kilo\electronvolt} electron arrives \SI{7.5}{\nano\second} earlier to the sample than a \SI{5}{\kilo\electronvolt} one, which corresponds to a delay stage movement of \SI{1.125}{\meter}. This delay line is also used to compensate for the different path lengths of the harmonics inside the harmonic generator, resulting in variations in their arrival time ($\sim\SI{1}{\nano\second}$). The physical length of DL1 is \SI{1.26}{\meter}.

In a pump-probe experiment, we tune the delay between electrons and light by moving the second delay line (DL2). We use a motorized linear stage (Newport M-IMS600BPP, and motion controller Newport ESP301-1G), with total range of \SI{60}{\centi\meter} (\SI{4}{\nano\second}), minimum step size of \SI{1.25}{\micro\meter} (\SI{8.3}{\femto\second}) and precision of \SI{0.65}{\micro\meter} (\SI{4}{\femto\second}). The stage movement is controlled using a script developed for the Odemis software (Delmic), such that its movement is integrated with the acquisition of optical data. In an experiment we typically choose the center of this delay line to correspond to the zero delay between electrons and light, meaning that we can scan in a $-2$ to \SI{+2}{\nano\second} range (with sign defined as the arrival time of the laser with respect to the electron). The temporal alignment of electron and laser beams on the sample is discussed in section \ref{sec:Setup-chapter_TimeAlignment}.

After DL2, the laser beam is directed towards the SEM chamber using an 8:92 pellicle beam splitter (Thorlabs BP208) or a dichroic mirror optimized for either the \nth{2} or the \nth{3} harmonic (Semrock Di02-R532-25x36 and Di01-R355-25x36, respectively). The position and angle of the beam splitter or dichroic mirror is controlled by a kinematic mount and a linear stage, thus allowing us to precisely align the laser with respect to the parabolic mirror. Finally, the light is focused on the sample using the parabolic mirror described in section \ref{sec:Setup-chapter_Overview}. The alignment of the laser beam on the sample is discussed in section \ref{sec:Setup-chapter_SpatialAlignment}.

\begin{figure}
	\centering
	\includegraphics[width=0.5\textwidth]{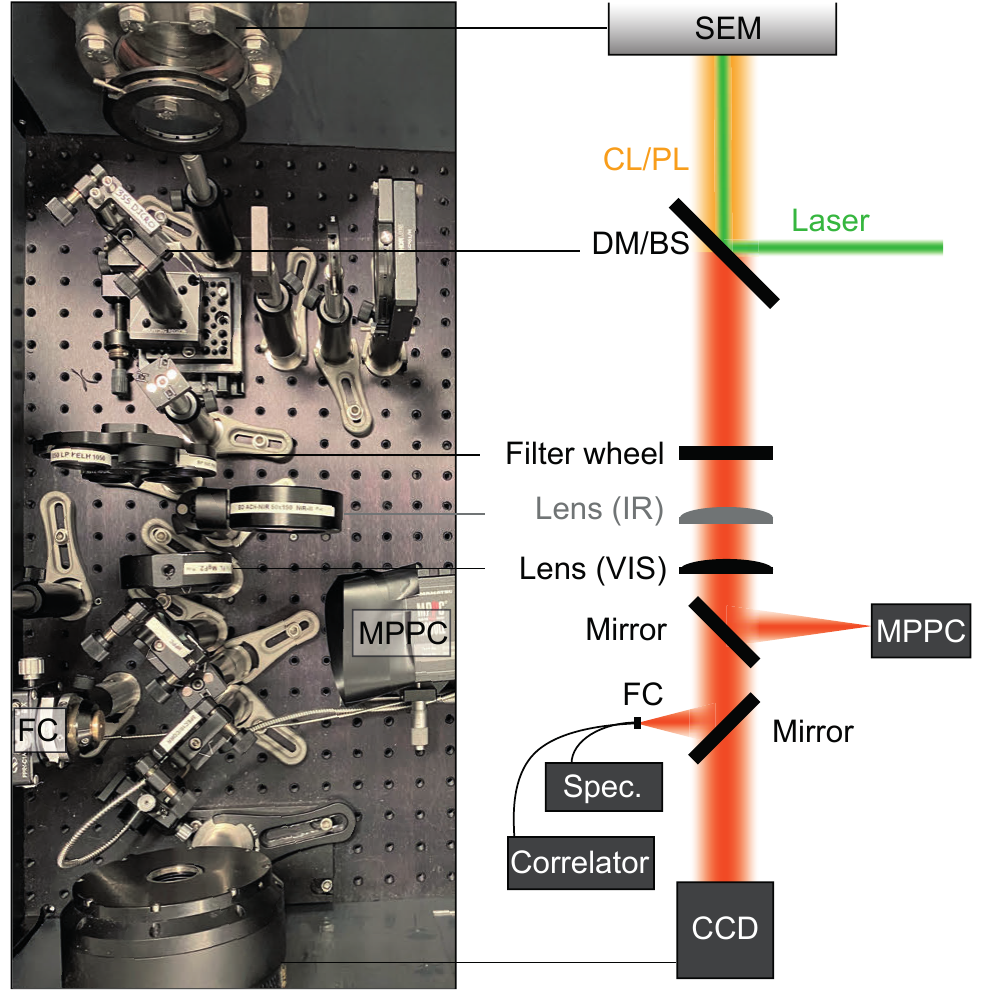}
	\caption{Luminescence (CL/PL) collection path. The light collected by the parabolic mirror is directed outside the SEM chamber for analysis. Different optical paths are available depending on the type of measurement: angular, spectroscopic, time-correlated or phase-locked. DM: dichroic mirror, BS: beam splitter, FC: fiber coupler, MPPC: multipixel photon counter, Spec: spectrometer.}
	\label{fig:Setup-chapter_CLpath}
\end{figure}

\section{Luminescence (CL/PL) collection path}
After excitation with an electron or laser beam (or both), the luminescence is collected by the parabolic mirror (described in section \ref{sec:Setup-chapter_Overview}). The resulting luminescence beam is collimated and has a size determined by the mirror dimensions (\SI{23x11}{\milli\meter}) \citep{Coenen2014,Brenny2016a}. The luminescence is further directed outside of the SEM towards the detection path setup. A photograph of this optical setup is provided in Fig. \ref{fig:Setup-chapter_CLpath}, together with the corresponding schematic. We have four types of detection methods: angular, spectral, time-correlated and phase-locked. The optical components in the collection path are either on magnetic mounts or can be easily removed, such that we have full flexibility for different optical configurations, depending on the experiment. In the next sections we describe each of these detection schemes.

\subsection{Alignment with CCD camera} \label{sec:Setup-chapter_CCD}
\begin{figure}
	\centering
	\includegraphics[width=0.5\textwidth]{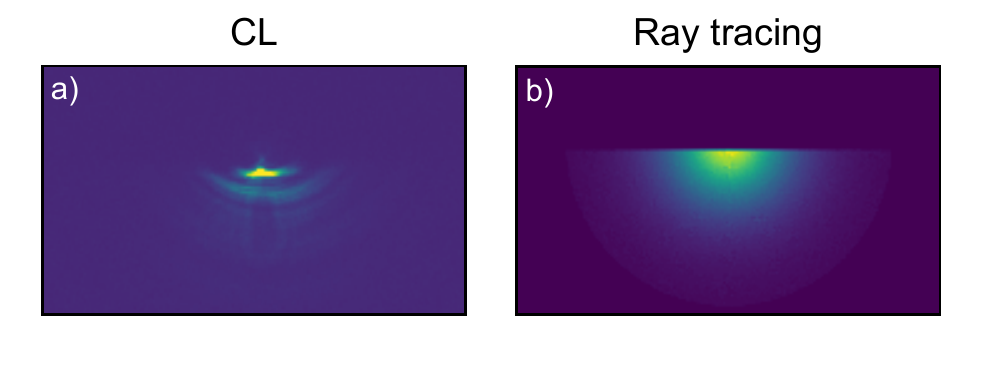}
	\caption{CL alignment. (a) Pattern of the collimated CL beam on the CCD obtained when the parabolic mirror is aligned with respect to the sample. The image was obtained when exciting a GaN sample with a \SI{5}{\kilo\electronvolt} continuous electron beam. (b) Ray tracing calculation of the image on the CCD obtained for a point source placed at the focal point of the parabolic mirror.}
	\label{fig:Setup-chapter_CLAlignment}
\end{figure}

The alignment of the parabolic mirror and sample height is performed by sending the CL light directly to a 2D back-illuminated thermoelectrically-cooled CCD silicon array (PI PIXIS 1024B, 1024x1024 pixels), operating at a temperature of \SI{-70}{\degreeCelsius}. In this case we do not place any additional optical components in the detection path, such that we directly collect the collimated CL beam. An example of the image obtained on the CCD for an aligned parabolic mirror and sample is provided in Figure \ref{fig:Setup-chapter_CLAlignment}a, together with a ray tracing calculation of the pattern on the CCD obtained for a point source placed at the focal point of the mirror (Fig. \ref{fig:Setup-chapter_CLAlignment}b) \citep{Vesseur2011,Coenen2014}.

\subsection{Spectroscopy} \label{sec:Setup-chapter_Spectroscopy}
Spectrally-resolved measurements in the visible range are performed by sending the emitted light through an achromatic lens ($f=\SI{160}{\milli\meter}$, $d=\SI{40}{\milli\meter}$) and a silver mirror (Thorlabs PF10-03-F01) to couple it to a multimode fiber with \SI{550}{\micro\meter} core diameter (OZ Optics QMMJ-55-IRVIS-550/600-3AS-2). The fiber is mounted on a manually controlled 2D mechanical stage to optimize the coupling of light into the fiber in the plane perpendicular to the optical axis. The fiber guides the light to a spectrometer (PI Acton SP2300i) containing a liquid-nitrogen-cooled silicon CCD array (PI Spec-10 100F/LN, 1340x100 pixels), which reaches a temperature of \SI{-120}{\degreeCelsius} for enhanced signal-to-noise ratio (SNR). 

\begin{figure}
	\centering
	\includegraphics[width=0.5\textwidth]{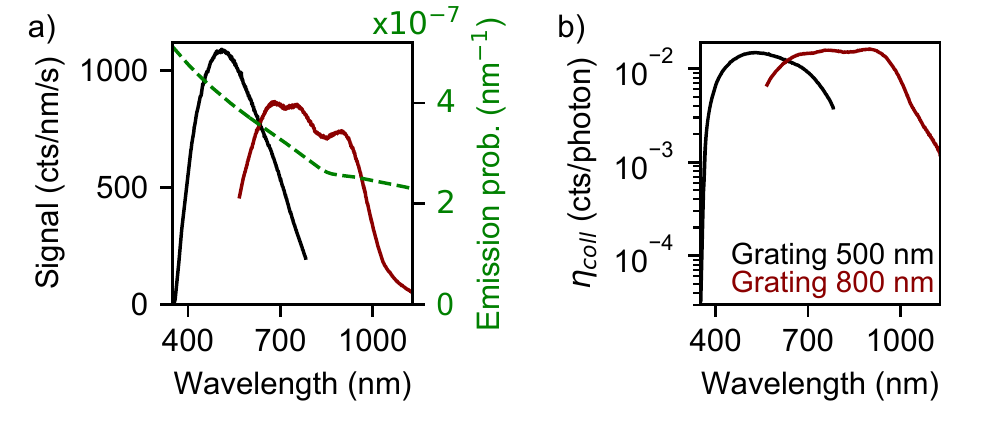}
	\caption{System response of spectral measurements. (a) Spectra of transition radiation of a single-crystal Al sample obtained under \SI{30}{\kilo\electronvolt} electron excitation (\SI{143.9}{\nano\ampere}) for a 150 gr/nm grating with 500 (black) and \SI{800}{\nano\meter} (dark red) blaze. Right axis (green): Theoretical TR emission probability per electron. (b) Collection efficiency of the setup using both gratings of the spectrometer.}
	\label{fig:Setup-chapter_TRAl}
\end{figure}

The system response of the spectral measurements is characterized by measuring the spectrum of transition radiation (TR) of a single-crystal Al sample, similar to previous works \citep{Brenny2014, Meuret2017}. Figure \ref{fig:Setup-chapter_TRAl}a shows TR spectra obtained upon excitation with a \SI{30}{\kilo\electronvolt} continuous electron beam (\SI{143.9}{\nano\ampere}) when using a 150 gr/nm grating with 500 and \SI{800}{\nano\meter} blaze (black and dark red, respectively). Fig. \ref{fig:Setup-chapter_TRAl}a also displays the calculated probability of photon emission per electron and wavelength bandwidth (green dashed curve), obtained using the formalism described in section IVC of ref. \citep{GarciaDeAbajo2010}. Using both curves we can extract the collection efficiency of our system ($\eta_{\textrm{collection}}$), defined as the number of counts detected per photon emitted by the sample. The results are shown in Fig. \ref{fig:Setup-chapter_TRAl}b for both gratings. The collection efficiency is not sample dependent, and can be used for any kind of CL and PL experiment, as long as we keep the same acquisition settings as in the TR measurements. Differences in the alignment of the parabolic mirror and fiber coupling can lead to changes of $\sim\SI{30}{\percent}$ in the collection efficiency \citep{Brenny2014}.

\subsection{Time-correlated measurements}\label{sec:Setup-chapter_TimeCorrelated}
We study the dynamics of laser or electron excitation and light emission using two types of time-correlated measurements: time-correlated single-photon counting (TCSPC) and second-order autocorrelation ($g^{(2)}(\tau)$) measurements.

\subsubsection*{Time-correlated single-photon counting}\label{sec:Setup-chapter_TCSPC}
\begin{figure}
	\centering
	\includegraphics[width=0.5\textwidth]{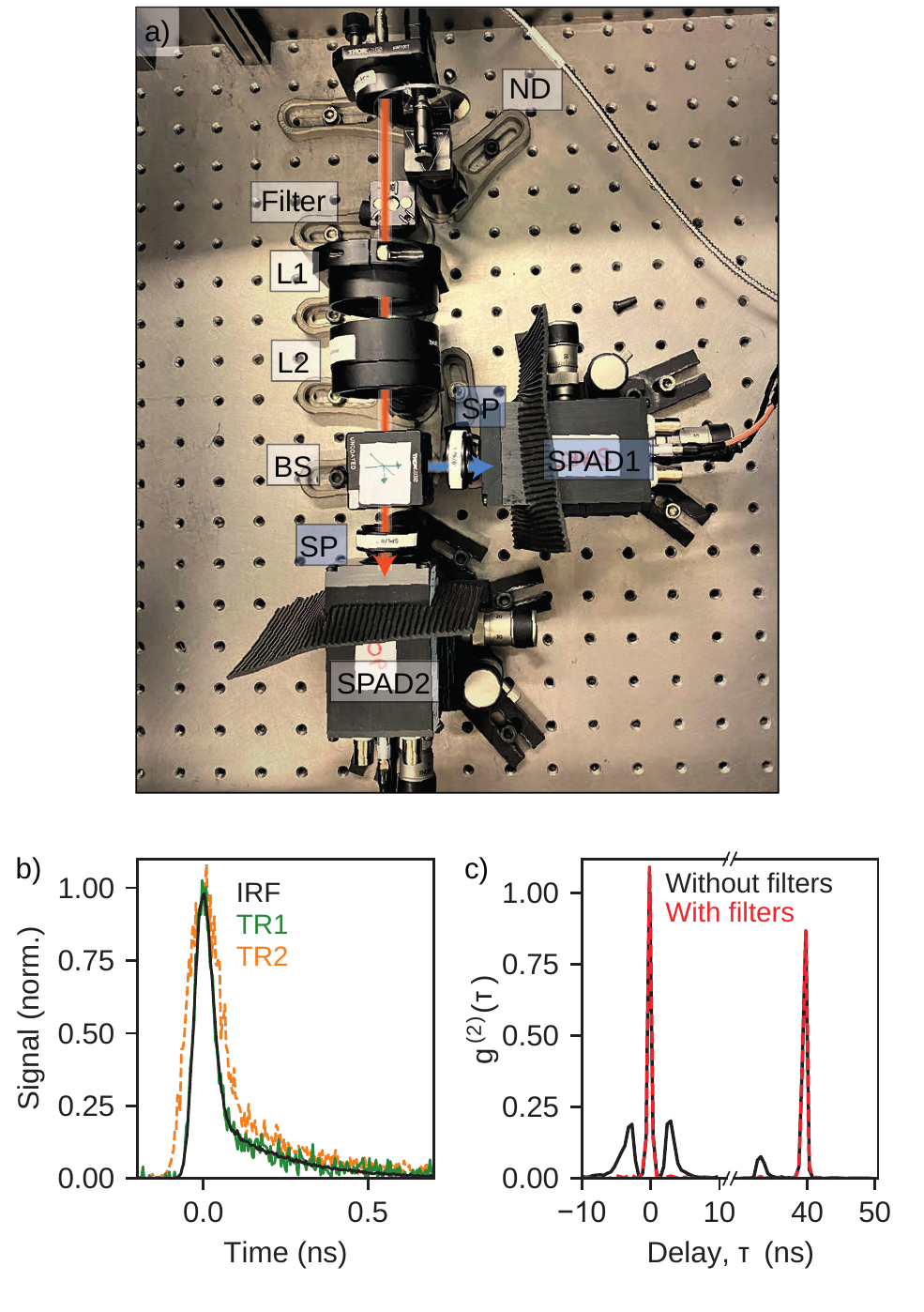}
	\caption{Time-correlated single-photon counting and autocorrelation $g^{(2)}(\tau)$ measurements. (a) Photograph of the optical setup used to perform time-correlated measurements. Here ND refers to a (tunable) neutral density filter, L1 and L2 are lenses, BS is a beam-splitter, SP is a short-pass filter and SPAD1 and SPAD2 are single-photon avalanche photodiodes. The orange arrow indicates the optical path for TCSPC measurements, while the blue arrow refers to the additional path to perform $g^{(2)}(\tau)$ measurements. (b) Time resolution of the TCSPC system ($\sim\SI{60}{\pico\second}$, black curve) obtained when sending the \nth{2} harmonic of the fs-laser towards the SPAD. The dispersion of the fiber is evaluated by measuring the time statistics of transition radiation on a single-crystal Au sample in the $620\pm\SI{5}{\nano\meter}$ (TR1, green curve) and $532-\SI{670}{\nano\meter}$ (TR2, orange curve) spectral ranges. (c) $g^{(2)}(\tau)$ measurements obtained when sending the \nth{2} harmonic beam of the fs laser to the Hanbury Brown and Twiss setup. The curve obtained without using any additional filter exhibits crosstalk (solid black). In contrast, the dashed red curve shows a measurement in which a \SI{670}{\nano\meter} shortpass and a $550\pm\SI{40}{\nano\meter}$ bandpass filters were used, thus reducing the effect of crosstalk.}
	\label{fig:Setup-chapter_Correlator}
\end{figure}

Time-correlated single-photon counting (TCSPC) measurements rely on the excitation of the sample with a pulsed beam (either electron or laser), and the subsequent analysis of the temporal statistics of the emitted light. We use the same optics (lens and mirror) as for the spectral measurements to couple the luminescence to a multimode fiber with \SI{105}{\micro\meter} core diameter (Thorlabs FG105UCA). The fiber is connected to an external optical setup, mounted on a portable breadboard. A photograph of the setup is shown in Fig. \ref{fig:Setup-chapter_Correlator}a, together with a schematic of the optical path (in orange). The luminescence is initially attenuated using a tunable neutral density filter and can be spectrally filtered with an optical filter. Next, the light goes through a set of lenses (both $f=\SI{7.5}{\centi\meter}$) to collimate and refocus it onto a single-photon avalanche photodiode (SPAD) (PicoQuant PDM Series). The SPAD has an active area of \SI{100x100}{\square\micro\meter} and is mounted on a 3D mechanical stage for optimum alignment with respect to the light beam. The SPAD has a detector dead time of $\sim\SI{100}{\micro\second}$, thus count rates above $\sim 10^6\SI{}{\per\second}$ would lead to a lower photon detection efficiency and detector damage. The entire correlator path is enclosed inside a light-tight enclosure in order to reduce background signal and protect the SPAD. 

TCSPC measurements are performed by recording a histogram of the photon arrival time with respect to a reference signal (trigger), which is done using a time-correlator (PicoQuant PicoHarp 300). We direct part of the laser beam (usually the \nth{2} or \nth{3} harmonic) to a photodiode (PicoQuant TDA 200), which sends an electrical signal to the time-correlator, thus acting as a trigger. The time-correlator calculates the difference in arrival time between the trigger pulse and the electric pulse generated by the SPAD after detection of a CL/PL photon, thus resulting in a histogram of photon arrival time. In TCSPC measurements, only the first photon that reaches the detector from each luminescence pulse is recorded. Hence, it is important to keep a low count rate, typically $<\SI{1}{\percent}$ of the laser rate, to avoid an overestimation of the number of photons collected in the first time bins (pile-up effect), which would lead to the recording of artificially fast dynamics \citep{Wahl2017}. To avoid this pile-up effect, the intensity of the light is controlled with a tunable neutral density filter such that, on average, less than one photon per pulse reaches the detector. 

The time resolution of the TCSPC setup is determined by the instrument response function (IRF) and dispersion in the optical fiber. We measure the IRF by directly sending the \nth{2} harmonic ($\lambda=\SI{517}{\nano\meter}$) of the fs-laser into the TCSPC setup, as shown in Fig. \ref{fig:Setup-chapter_Correlator}b (black curve). Considering that the laser pulse width is negligible ($\sim\SI{250}{\femto\second}$), we obtain an IRF of $\SI{60}{\pico\second}$ (FWHM), which is determined by the precision of the correlator, SPAD and triggering photodiode \citep{Meuret2019}. In the case of a spectrally broad luminescence signal, we also need to account for dispersion in the optical fiber. In order to quantify this effect, we measured CL time traces of transition radiation (TR) on a single-crystal Au sample using different optical filters, given that TR emission is spectrally broadband (see Fig. \ref{fig:Setup-chapter_TRAl}a). Fig. \ref{fig:Setup-chapter_Correlator}b shows traces obtained when filtering the signal in the $620\pm\SI{5}{\nano\meter}$ (TR1, green) and $532-\SI{670}{\nano\meter}$ (TR2, orange) spectral ranges. In this case we used a \SI{30}{\kilo\electronvolt} pulsed electron beam containing an average of $40\pm15$ and $80\pm30$ electrons per pulse (green and orange curves, respectively) ($V_\textrm{ext}=\SI{650}{\volt}, C_1=\SI{1050}{\volt}$ \citep{Meuret2019,Sola-Garcia2021a}). TR emission can be assumed as instantaneous ($\sim\SI{20}{\femto\second}$ \citep{Brenny2016}), hence the width of the time trace is determined by the IRF, dispersion of the fiber and electron pulse width ($\sim\SI{}{\pico\second}$ \citep{Sun2016a,Meuret2019}). We observe that the curve obtained when filtering the luminescence in the $620\pm\SI{5}{\nano\meter}$ range resembles the one for the IRF obtained with the laser, thus suggesting that both dispersion in the fiber and electron pulse width are negligible. Instead, measuring luminescence in a broader range ($532-\SI{670}{\nano\meter}$) results in a broader time trace, which we attribute to dispersion in the optical fiber. From the measurements we estimate a dispersion of $\sim\SI{0.2}{\pico\second\per\nano\meter\per\meter}$, which is reasonable for a glass-type fiber. The temporal broadening due to dispersion in the fiber could be removed by having a completely free-space coupling system.

\subsubsection*{Second-order autocorrelation ($g^{(2)}(\tau)$) measurements}
To gain further insights in laser and electron excitation dynamics, we can study the CL and PL photon correlation statistics, which are measured using second-order autocorrelation ($g^{(2)}(\tau)$) measurements. Given a time-dependent luminescent intensity $I(t)$, $g^{(2)}(\tau)$ is defined as \citep{Fox2006}
\begin{equation}
	g^{(2)}(\tau)=\frac{\left\langle I(t)I(t+\tau)\right\rangle}{\langle I(t)\rangle^2}\:,
\end{equation}
where the angle brackets denote the time average. Hence, in a $g^{(2)}(\tau)$ measurement we build a histogram of the number of coincidence events, defined as the detection of two photons, with respect to the time delay between them ($\tau$). Our $g^{(2)}(\tau)$ experiments are performed using a Hanbury Brown and Twiss geometry \citep{HanburyBrown1956a}. We use the same optical setup as for TCSPC measurements, with the difference that we now use two SPADs (1 and 2 in Fig. \ref{fig:Setup-chapter_Correlator}a, blue path). Both detectors are connected to the time-correlator. A 50:50 beam splitter is placed after the last lens, such that the CL/PL photons have equal probability of being detected by each SPAD. After detection of a photon by SPAD1 at a given time $t_1$, the time-correlator acts as a stopwatch until a photon is detected on SPAD2 (at a time $t_2$). A count is added on the histogram at a delay corresponding to $\tau=t_2-t_1$. A $g^{(2)}(\tau)$ measurement is always symmetric, given that there is an equal probability of detecting a photon first on SPAD1 and then SPAD2 ($+\tau$) or viceversa ($-\tau$). An example of a $g^{(2)}(\tau)$ measurement is shown in Fig. \ref{fig:Setup-chapter_Correlator}c (dashed red), which was obtained when directing the \nth{2} harmonic of the laser beam to the Hanbury Brown and Twiss setup. The $g^{(2)}(\tau)$ curve exhibits peaks separated by the time between pulses (\SI{39.7}{\nano\second} at \SI{25.19}{\mega\hertz}), corresponding to correlations between photons from the same pulse ($\tau=0$), from the next pulse ($\tau=\SI{39.7}{\nano\second}$) and so on \citep{Sola-Garcia2021}. 

A potential concern in $g^{(2)}(\tau)$ measurements is the phenomenon of crosstalk. After detection of a photon, SPADs can emit secondary photons, typically in the infrared spectral range \citep{Rech2008}. In $g^{(2)}(\tau)$ measurements, the detection of this secondary photon by the second SPAD leads to the appearance of peaks at specific delays in the $g^{(2)}(\tau)$ curve. Figure \ref{fig:Setup-chapter_Correlator}c (solid black) shows a $g^{(2)}(\tau)$ measurement obtained when sending the \nth{2} harmonic of the laser beam to the Hanbury Brown and Twiss setup. We observe a peak at $\tau=0$, corresponding to the correlations between photons from the same pulse. We also observe two additional peaks separated by $\SI{2.6}{\nano\second}$ from the main peak, corresponding to crosstalk. Using shortpass filters allows us to filter out most of the secondary photons, thus reducing the crosstalk. Figure \ref{fig:Setup-chapter_Correlator}c (dashed red) shows a measurement obtained when using a \SI{670}{\nano\meter} shortpass filter in front of each SPAD and an additional bandpass filter ($550\pm\SI{40}{\nano\meter}$) right after the fiber output. In this case, the two peaks corresponding to crosstalk are almost completely suppressed. Hence, in this setup, $g^{(2)}(\tau)$ measurements are limited to low wavelengths ranges, below $\sim\SI{600}{\nano\meter}$.

\subsection{Lock-in detection} \label{sec:Setup-chapter_LockIn}
The previously described detection systems are based on the direct collection and analysis of the emitted light. In pump-probe measurements we have two signals: photoluminescence and cathodoluminescence, usually with very different magnitudes. This means that any spectrally or temporally-resolved measurement performed with the methods described above will be dominated by the largest signal. Hence, the analysis of the weaker signal becomes challenging, given that it can become buried in the noise of the larger signal. Moreover, in many samples CL and PL emission exhibit a similar spectrum, thus making it difficult to spectrally filter out one of those. In our PP measurements, PL is typically several orders of magnitude larger than the CL signal, partially due to the larger PL spot area compared to the CL one, as will be discussed below (section \ref{sec:Setup-chapter_LaserSpot}). Moreover, when using the laser as a pump, a large excitation fluence is usually needed to achieve nonlinear regimes, thus further increasing the PL/CL ratio.

A method to extract the weaker signal (here, CL) is by decreasing the measurement bandwidth, such that noise is reduced, by means of lock-in detection. In this case, we focus the luminescence (CL/PL) on a thermoelectrically-cooled multi-pixel photon counter (MPPC module, Hamamatsu C14455-3050GA, 2836 pixels). The MPPC is connected to a lock-in amplifier (Stanford Research Systems RS830 DSP). Lock-in amplifiers work as phase-sensitive detectors, which can isolate small signals modulated at a known frequency and filter out other frequency components, thus improving the signal-to-noise ratio. This allows us to separate the desired signal from noise or background signal. In our experiments, we use an optical chopper (Thorlabs MC2000B-EC) to modulate the laser beam that we use to generate electron pulses (\nth{4} harmonic), typically at a few hundreds of \SI{}{\hertz}, thus resulting in a modulated CL emission. The chopper is also connected to the lock-in amplifier and serves as a reference signal. This mechanism allows us to isolate the CL signal from a large PL background, given the difference in CL and PL modulations.

In a pump-probe measurement we usually compare the magnitude of the probe signal (either CL or PL) with and without pumping the sample (with laser or electrons, respectively). Hence, the detector (MPPC in our case) should have a low noise level so that it can detect small signals, in our case CL ($<\SI{}{\pico\joule}$), and a large dynamic range so that large PL signals do not saturate the detector, thus causing nonlinear behavior. Saturation of the MPPC during the experiment would result in artificially low signals in the pump-probe measurement compared to the reference (only CL or PL) one.

\begin{figure}
	\centering
	\includegraphics[width=0.5\textwidth]{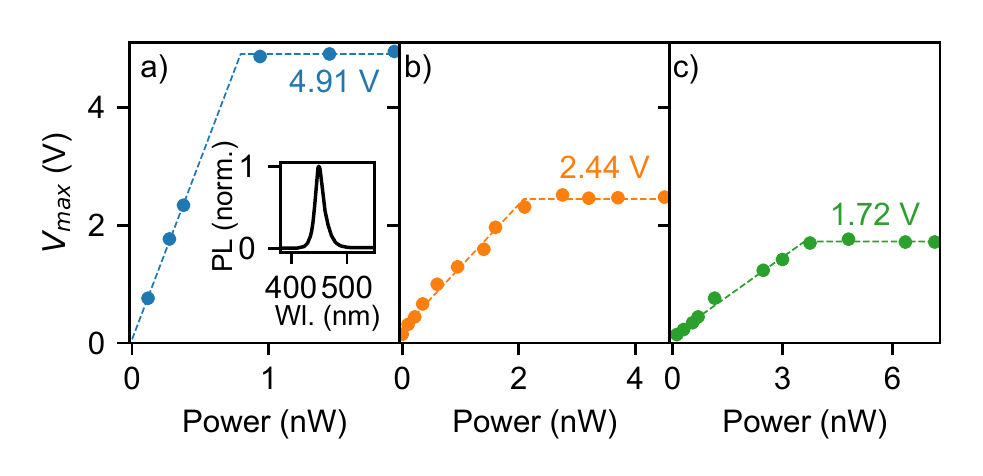}
	\caption{Response curve of the multi-pixel photon counter (MPPC) module. Dependence of the output voltage on the PL power incident on the MPPC at 1.00, 5.04 and \SI{25.19}{\mega\hertz} repetition rates (a,b and c, respectively). The measurement was performed when exciting InGaN/GaN quantum wells with the \nth{3} harmonic of the laser. The inset in (a) shows the PL spectrum. }
	\label{fig:Setup-chapter_MPPC}
\end{figure}

The output voltage of the MPPC as a function of incident power is shown in Figure \ref{fig:Setup-chapter_MPPC} for different laser repetition rates. Here the input power is PL emission from InGaN/GaN quantum wells upon excitation with a $\lambda=\SI{345}{\nano\meter}$ laser beam (\nth{3} harmonic). The PL spectrum is shown in the inset of Fig. \ref{fig:Setup-chapter_MPPC}a, and shows emission in the $400-\SI{450}{\nano\meter}$ spectral range \citep{Meuret2017,Sola-Garcia2021}. We chose to perform the characterization using PL emission, instead of directly sending the laser towards the MPPC, to mimic the conditions of an actual pump-probe experiment. The PL power was measured using a silicon photodiode (Thorlabs S120VC) placed along the PL path. Given the lower sensitivity of the photodiode compared to the MPPC, we placed a neutral density (ND2) filter in front of the MPPC to attenuate the incoming PL. The MPPC shows a linear response up to a certain power $P_\textrm{max}$, above which saturation is observed. The curves obtained at different repetition rates exhibit a different $P_\textrm{max}$, ranging from \SI{3.6}{\nano\watt} at \SI{25.19}{\mega\hertz} down to \SI{0.8}{\nano\watt} at \SI{1.00}{\mega\hertz}. We obtain a maximum output voltage of \SI{4.98}{\volt} at \SI{1.00}{\mega\hertz}, very close to the expected maximum output of the MPPC (\SI{5}{\volt}). Instead, the maximum voltage goes down to $2.44$ and \SI{1.72}{\volt} for $5.04$ and \SI{25.19}{\mega\hertz}, respectively. The dependence of the maximum power and output voltage on the repetition rate is attributed to the pixel recovery time. This analysis shows that in a pump-probe experiment it is crucial to ascertain that the output voltage of the signals (CL, PL and CL+PL) is sufficiently below the maximum output at the operating repetition rate.

\section{Laser focusing on the sample}\label{sec:Setup-chapter_SpatialAlignment}
We use the same parabolic mirror for luminescence collection (section \ref{sec:Setup-chapter_Overview}) to focus the laser beam on the sample. This configuration allows us to focus the light down to a micrometer size spot without introducing additional components in the SEM chamber. Focusing with a parabolic mirror requires a precise alignment, given that any small misalignment can lead to aberrations, thus degrading the shape of the laser spot on the sample \citep{Howard1979, Drechsler2001}. Moreover, a precise spatial overlap of electrons and light is essential in pump-probe experiments. In this section we investigate the spot size of the laser on the sample and its alignment with respect to the electron beam.

In the experiments, we align the parabolic mirror by bringing the sample into focus while optimizing the CL pattern on a CCD camera, as explained in section \ref{sec:Setup-chapter_CCD}. This guarantees maximum collection of the CL emitted light. The laser beam is then aligned on the sample by mechanically tuning the angle of the dichroic mirror or beam-splitter (see Fig. \ref{fig:Setup-chapter_LaserAlignment}a) and position it with respect to the parabolic mirror. We can also use the feedback system to precisely tune the mirror actuators, thus yielding a higher angular and spatial control.

\subsection{Characterization of the laser focus} \label{sec:Setup-chapter_LaserSpot}
\begin{figure}
	\centering
	\includegraphics[width=0.5\textwidth]{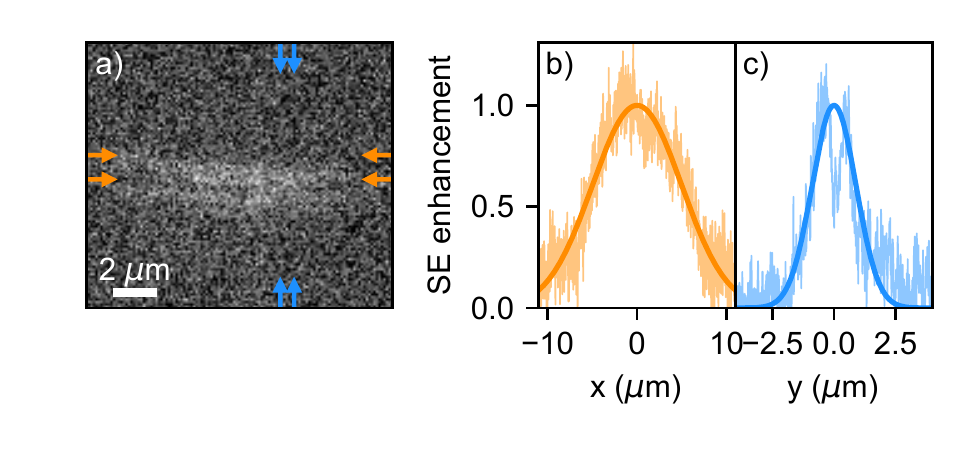}
	\caption{Visualization of the laser spot on the sample. (a) SE image of a GaN substrate obtained after excitation with the \nth{3} harmonic laser beam. Horizontal (b) and vertical (c) cross-sections of the SE image together with the corresponding Gaussian fits, from which we obtain a laser spot size of 11.7 and \SI{2.1}{\micro\meter} in the horizontal and vertical direction, respectively.}
	\label{fig:Setup-chapter_LaserSpot}
\end{figure}

Direct imaging of the laser spot on the sample plane is challenging due to its size (\SI{}{\micro\meter}) and limited space in the SEM chamber. Instead, we can examine the change in secondary electron (SE) yield after optical excitation. Previous works have studied changes in SE emission in a pump-probe configuration \citep{Liao2017a, Garming2020}, from which the laser spot was imaged on the sample. For simplicity, here we rely on non-reversible changes in the SE contrast induced after repeated excitation with the laser (typically >10 s). Figure \ref{fig:Setup-chapter_LaserSpot}a shows an SE image of a \SI{30}{\micro\meter}-thick GaN film on a sapphire substrate (PI-KEM, undoped n-type) after \SI{10}{\second} exposure with the \nth{3} harmonic laser beam ($\lambda=\SI{345}{\nano\meter}$, \SI{1.7}{\milli\watt} average power at \SI{5.04}{\mega\hertz}). The scan was taken using a \SI{5}{\kilo\electronvolt} electron beam with \SI{380}{\pico\ampere} electron current and \SI{10}{\micro\second} pixel integration time. We observe a centered elongated spot with a higher SE yield, which we attribute to the laser-exposed area. This is further confirmed by tracking the movement of this spot as we misalign the laser beam with respect to the sample, as will be explained below. The mechanism behind the change in SE signal upon laser excitation is unknown. We use a laser fluence ($\sim \SI{0.4}{\milli\joule\per\square\centi\meter}$) well below the reported damage threshold of GaN under UV fs-laser excitation ($\sim \SI{5}{\joule\per\square\centi\meter}$) \citep{Eliseev1999}. Hence, it is unlikely that laser ablation plays a role. Previous studies have reported a reduction of the surface roughness of GaN upon excitation with UV ns laser pulses \citep{Akane1999,Dubowski2001}, which could explain the change in SE yield. The change in SE contrast could also be due to increased contamination on the optically-excited surface, similar to the contamination typically observed in SEMs in the electron-irradiated areas \citep{Vladar2005,Griffiths2010}. Further experiments with varying laser power, repetition rate and exposure time could be performed to elucidate the origin of this effect.

We further analyze the shape of the laser spot by taking horizontal ($x$) and vertical ($y$) cross-sections of the SE image, as shown in Fig. \ref{fig:Setup-chapter_LaserSpot}b and c, respectively. In each case, the curve is obtained by integrating over the rectangle delimited by the corresponding arrows (orange and blue in Fig. \ref{fig:Setup-chapter_LaserSpot}a). The position of this rectangle is chosen in both cases such that it yields the largest spread, thus ensuring that we characterize the largest section of the laser spot. In this experiment we ensured that the laser power was low enough to avoid any saturation effect. The solid lines represent Gaussian fits, from which we derive a laser spot size (full width at half maximum) of $11.7$  and \SI{2.1}{\micro\meter} in the horizontal and vertical directions, respectively. We attribute the asymmetry of the laser spot to non-perfect alignment of the laser beam with respect to the parabolic mirror. 

\subsection{Laser alignment on the sample} \label{sec:Setup-chapter_LaserAlignment}
As shown above, imaging the change in SE yield after laser excitation allows us to characterize the shape and size of the laser spot on the sample. However, we have only observed this effect on specific samples and under excitation with the \nth{3} harmonic laser beam. Hence, it is not practical for alignment in regular pump-probe experiments. Instead, we can align the laser beam on the sample by analyzing the image on the CCD of the PL beam, similar to the method used to align the parabolic mirror and sample height with CL (section \ref{sec:Setup-chapter_CCD}).

\begin{figure*}
	\centering
	\includegraphics[width=0.95\textwidth]{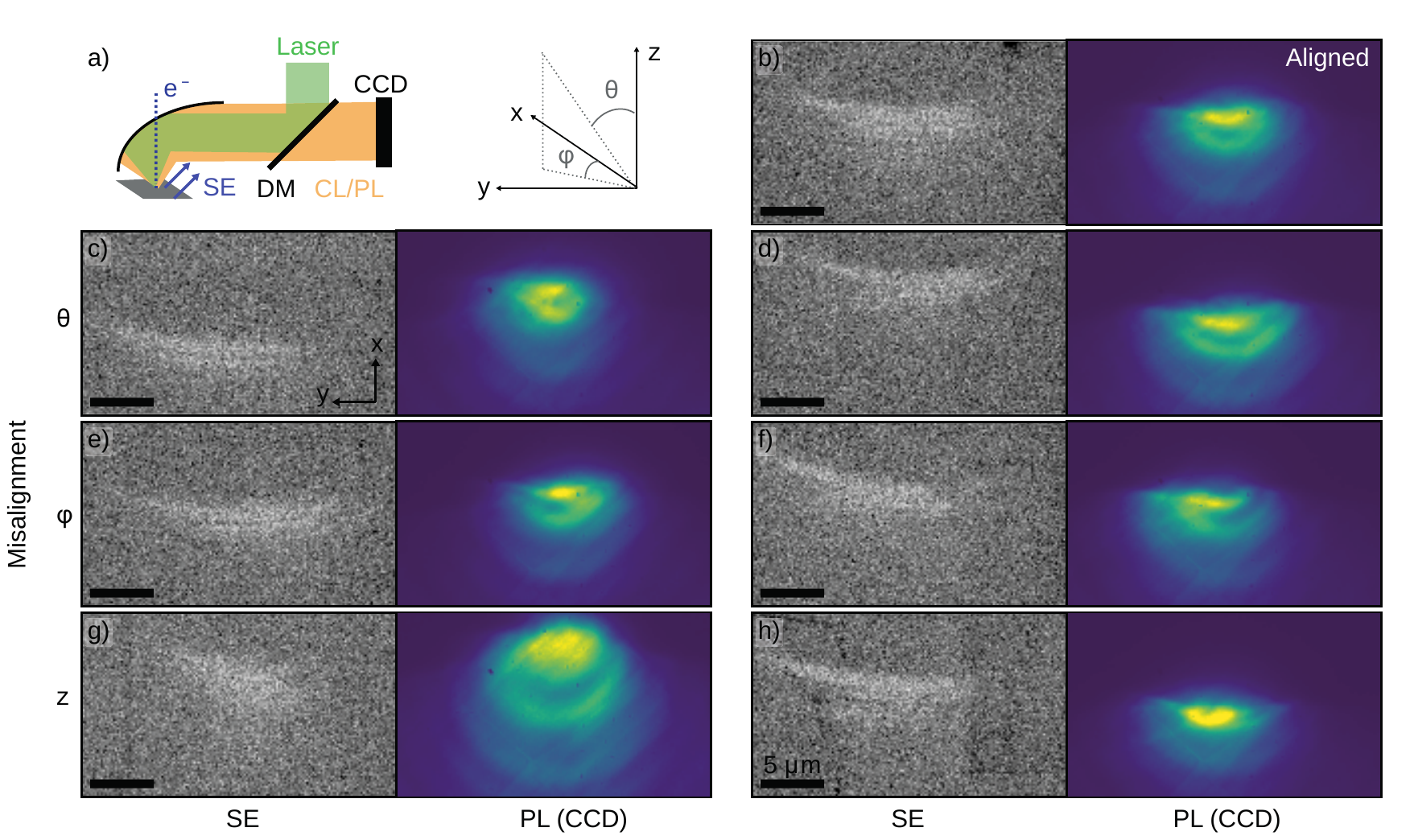}
	\caption{Laser alignment on the sample. (a) Left: schematic of the experiment, including PL collection on the CCD camera and secondary electron (SE) image formation. The position of the laser on the sample is tuned by moving the dichroic mirror (or beam splitter) (DM). Right: system of coordinates. (b-h) SE images of the GaN surface after excitation with the \nth{3} harmonic laser beam (left) together with the corresponding PL pattern on the CCD (right) for an aligned laser beam (b) and misalignments in the polar ($\theta$) (c,d) and azimuthal ($\phi$) (e,f) directions, as well as in the sample height (g,h).}
	\label{fig:Setup-chapter_LaserAlignment}
\end{figure*}

We define good alignment of the laser beam as when it is centered with respect to the electron beam, that is, when it is at the focal point of the parabolic mirror. Figures \ref{fig:Setup-chapter_LaserAlignment}(b-h) show various images of PL emission from GaN upon laser excitation (\nth{3} harmonic beam, $\lambda=\SI{345}{\nano\meter}$), together with the corresponding SE image of the laser spot on the sample, obtained using the method discussed above. Panel (b) shows an example of satisfactory alignment of the laser. Here the CCD image resembles the one expected for a collimated beam from an off-axis parabolic mirror, as shown in Fig. \ref{fig:Setup-chapter_CLAlignment}, and we observe that the laser spot is centered with respect to the electron beam. The rest of panels in Fig. \ref{fig:Setup-chapter_LaserAlignment} show the CCD pattern and SE image for different types of misalignments of the incoming angle of the laser (polar, $\theta$, and azimuthal, $\phi$, angles) (c-f) and height of the sample (g, h). As a reference, a schematic of the experiment and the coordinate system is shown in Fig. \ref{fig:Setup-chapter_LaserAlignment}a. The angle of the laser with respect to the parabolic mirror is tuned by controlling the tilt of the dichroic mirror (DM). We observe a clear correlation of the misalignment of the laser spot with respect to the electron beam and the pattern on the CCD. This shows that we can rely on the alignment of the laser using this technique, instead of having to image the laser spot, as in the previous section.

We should note that the height of the sample is fixed through the CL alignment, corresponding to the optimum collection of CL. This alignment should also correspond to best focus of the laser for a perfectly-collimated laser beam. Imperfections in the collimation can result in a focal point slightly different than the one for CL collection.

\section{Temporal alignment}\label{sec:Setup-chapter_TimeAlignment}
\begin{figure}[h]
	\centering
	\includegraphics[width=0.5\textwidth]{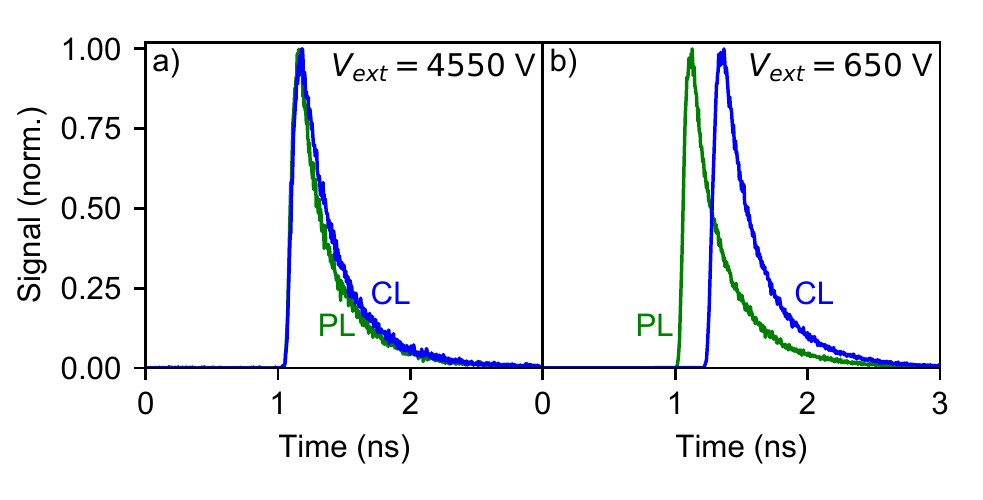}
	\caption{Temporal alignment of electron and light pulses. (a) PL (green) and CL (blue) decay traces from GaAs showing a good temporal overlap between electron and light pulses exciting the sample. (b) Same as in (a) but after lowering the extractor voltage, which results in a change in the arrival time of electrons.}
	\label{fig:Setup-chapter_TimeAlignment}
\end{figure}

Pump-probe experiments require precise control of the timing between pump and probe (electron and laser, or vice versa). Here we describe the temporal alignment of the laser and electron pulses. We record the decay statistics of PL and CL separately using the TCSPC setup (see \ref{sec:Setup-chapter_TCSPC}), from which we obtain the difference in arrival time between electron and laser pulses on the sample. Figure \ref{fig:Setup-chapter_TimeAlignment} shows decay traces of GaAs bandgap emission upon excitation with the laser beam ($\lambda_\textrm{exc}=\SI{345}{\nano\meter}$, $\sim\SI{5}{\milli\watt}$) (green curve) and \SI{30}{\kilo\electronvolt} electron pulses ($\sim 15$ electrons per pulse) (blue curve). The extractor voltage at the electron gun was set to \SI{4550}{\volt}. The x-axis indicates the time at which photons are detected on the SPAD with respect to the trigger signal. The temporal overlap of both traces, shown in Fig. \ref{fig:Setup-chapter_TimeAlignment}a, indicates a good time alignment between electron and laser pulses. The accuracy of this method for determining the zero-delay is limited by the minimum bin size of the time-correlator (\SI{4}{\pico\second}) and uncertainty in the determination of the arrival time of electrons or light from the decay curves. The latter becomes more complex when PL and CL exhibit different decay dynamics. Hence, we typically achieve an accuracy of $\sim\SI{10}{\pico\second}$. A higher accuracy in the determination of the zero-delay between electron and laser pulses can be obtained directly through a pump-probe experiment. In that case, the precision of the delay line stage is $\sim\SI{8}{\femto\second}$ (see section \ref{sec:Setup-chapter_PLpath}), hence the temporal resolution is only limited by the electron ($\sim \SI{}{ps}$) and laser ($\sim\SI{250}{\femto\second}$) pulse duration, as will be discussed below.

We should note that small changes in the electron or light path directly impact the temporal alignment. As we discussed previously, changing the energy of the electrons from 5 to \SI{30}{\kilo\electronvolt} results in a delay of  \SI{7.5}{\nano\second} (section \ref{sec:Setup-chapter_PLpath}). The conditions in the photoemission of electrons, such as extractor voltage, also determine the arrival time of the electrons. Figure \ref{fig:Setup-chapter_TimeAlignment}b shows decay traces of GaAs obtained under the same conditions as in (a) but at low extractor voltage ($V_\textrm{ext}=\SI{650}{\volt}$). We observe that the CL is now delayed by $\sim\SI{200}{\pico\second}$ with respect to the PL, which is attributed to the different speed at which electrons travel along the extractor plate. Hence, a precise temporal alignment is essential before starting a pump-probe experiment. Such accurate measurements of the timing of electron arrival can give further insights in electron  dynamics in both the electron source and column, both in SEM and TEM.

\section{Other considerations in PP-CL experiments}
The sections above describe the technical aspects in the development of a PP-CL microscope. Here, we discuss some extra considerations that need to be taken into account when designing a PP-CL experiment.

\subsection{Temporal resolution}
One of the main advantages of using pump-probe techniques is exploiting their high time resolution. The latter only depends on the delay between the two beams and is thus independent of the temporal precision of detectors, as happens in other time-resolved techniques. Given that typical delay stages have \SI{}{\femto\second} precision, the time resolution is typically determined by the pulse duration of one (or both) beams. In PP-CL, the duration of the electron pulse right after emission is determined by the pulse length of the laser pulse and the dynamics of the photoemission process, usually considered negligible compared to the duration of the laser pulse (hundreds of \SI{}{\femto\second}). However, the photoemitted electrons exhibit an energy distribution, typically in the $\SI{0.3}{}-\SI{2}{\electronvolt}$ \citep{Sola-Garcia2021a}, which translates into a chirp, given that electrons traveling at different speeds arrive at a different time on the sample \citep{Park2012}. Therefore, the electron pulse duration on the sample is larger than the duration of the initial optical pulse and depends on the conditions of photoemission \citep{Sola-Garcia2021a}.

A way of minimizing this temporal spread of the electron pulse is by working in the regime of less than 1 electron per pulse, on average \citep{Cook2016}. In this case, Coulomb interactions between electrons from the same pulse, and the resulting energy spread, are minimized. This approach is usually taken in other PP-type techniques using electron beams \citep{Park2007,Feist2017a}. However, the low electron dose used in this regime can result in a very small CL signal, in the case in which the electron is used as a probe, or low disturbance of the sample, when using the electron as a pump.

A second limitation of the pulse duration of the electron beam is given by the energy spread and electron speed in the SEM \citep{Gahlmann2008}. Even in the single-electron regime, electrons have an intrinsic energy spread, typically between \SI{0.3}{} and \SI{1}{\electronvolt}, determined by the conditions of the photoemission process. At \SI{30}{\kilo\electronvolt} this energy spread translates into a temporal spread below a the ps regime, mostly limited by the laser pulse duration. Lowering the electron energy down to \SI{5}{\kilo\electronvolt} results in electron pulse durations of at least \SI{1}{\pico\second}, in the case of the lowest energy spread (\SI{0.3}{\electronvolt}), given the lower speed of electrons. Further information about the energy spread of electrons in an ultrafast SEM and their impact on the electron pulse duration is given in ref. \citep{Sola-Garcia2021a}.

\subsection{Spatial resolution}
A significant advantage of PP-CL measurements compared their fully optical counterparts is exploiting the high spatial resolution given by the electron beam. Similar to the temporal resolution, the best spatial resolution is achieved for beams containing a low number of electrons per pulse. In this case  Coulomb repulsion, which would lead both to transverse spread and to chromatic aberrations due to the Boersch effect, is minimized \citep{Feist2017a}. Moreover, higher spatial resolutions are obtained when using emission conditions that result in a low number of photoemitted currents \citep{Meuret2019}. However, working in the regime of low number of electrons per pulse can be challenging when studying samples with low quantum efficiency.

\subsection{Electron as a pump}
An additional advantage of PP-CL measurements with respect to other electron-light configurations is the fact that we can now use the electron pulses as a pump, while analyzing the change in PL. Other configurations combining electron and laser pulses rely on the analysis of the electron signal, either transmitted or secondary electrons, and thus the sample is always pumped with light. An example of an experiment using the electron as a pump is given in ref. \citep{Sola-Garcia2019} when studying nitrogen-vacancy (NV) centers in diamond. Here the sample was excited with electrons and the resulting charge state of the NV centers was probed with a laser beam, showing that electron excitation induces the conversion of centers from the NV$^-$ to the NV$^0$ state.


\section{Conclusion}
This paper provides a complete overview of the development of the first pump-probe cathodoluminescence microscope. We discuss the technical implementation of the setup, based on the coupling of a fs laser to the electron cathode and sample chamber of a Quanta 250 FEG SEM. We also review and characterize the different methods to analyze the luminescence emitted by the sample (either PL or CL), including spectroscopy, time-resolved and autocorrelation measurements, and lock-in based detection. Spatial alignment is performed by visualizing the laser spot on the sample, or through the analysis of the angular emission pattern of the PL light, while temporal overlap is achieved by monitoring the arrival time of electrons and light through time-resolved PL and CL measurements. Overall, this paper provides a detailed discussion and characterization of a PP-CL microscope which can be used as reference for further developments. We envision that PP-CL can be used as a new tool to understand ultrafast dynamics in materials at the nanoscale.

\section*{Acknowledgements}
We gratefully acknowledge Erik Kieft (Thermo Fisher) for useful discussions and advice regarding the ultrafast SEM. We also thank Dion Ursem and Hans Zeijlemaker for their technical support and advice in the development of the PP-CL setup. This work is part of the research program of AMOLF which is partly financed by the Dutch Research Council (NWO). This project has received funding from the European Research Council (ERC) under the European Union’s Horizon 2020 research and innovation program (grant agreement No. 695343, SCEON), and the FET-Proactive program (grant agreement No. 101017720, EBEAM). S.M. acknowledges support from the French ANR funding agency through the ANR-19-CE30-0008-ECHOMELO grant.

\bibliography{PP-CLmicroscopy}

\begin{thebibliography}{55}%
\makeatletter
\providecommand \@ifxundefined [1]{%
 \@ifx{#1\undefined}
}%
\providecommand \@ifnum [1]{%
 \ifnum #1\expandafter \@firstoftwo
 \else \expandafter \@secondoftwo
 \fi
}%
\providecommand \@ifx [1]{%
 \ifx #1\expandafter \@firstoftwo
 \else \expandafter \@secondoftwo
 \fi
}%
\providecommand \natexlab [1]{#1}%
\providecommand \textit  [1]{``#1''}%
\providecommand \bibnamefont  [1]{#1}%
\providecommand \bibfnamefont [1]{#1}%
\providecommand \citenamefont [1]{#1}%
\providecommand \href@noop [0]{\@secondoftwo}%
\providecommand \href [0]{\begingroup \@sanitize@url \@href}%
\providecommand \@href[1]{\@@startlink{#1}\@@href}%
\providecommand \@@href[1]{\endgroup#1\@@endlink}%
\providecommand \@sanitize@url [0]{\catcode `\\12\catcode `\$12\catcode
  `\&12\catcode `\#12\catcode `\^12\catcode `\_12\catcode `\%12\relax}%
\providecommand \@@startlink[1]{}%
\providecommand \@@endlink[0]{}%
\providecommand \url  [0]{\begingroup\@sanitize@url \@url }%
\providecommand \@url [1]{\endgroup\@href {#1}{\urlprefix }}%
\providecommand \urlprefix  [0]{URL }%
\providecommand \Eprint [0]{\href }%
\providecommand \doibase [0]{http://dx.doi.org/}%
\providecommand \selectlanguage [0]{\@gobble}%
\providecommand \bibinfo  [0]{\@secondoftwo}%
\providecommand \bibfield  [0]{\@secondoftwo}%
\providecommand \translation [1]{[#1]}%
\providecommand \BibitemOpen [0]{}%
\providecommand \bibitemStop [0]{}%
\providecommand \bibitemNoStop [0]{.\EOS\space}%
\providecommand \EOS [0]{\spacefactor3000\relax}%
\providecommand \BibitemShut  [1]{\csname bibitem#1\endcsname}%
\let\auto@bib@innerbib\@empty
\bibitem [{\citenamefont {Merano}\ \emph {et~al.}(2005)\citenamefont {Merano},
  \citenamefont {Sonderegger}, \citenamefont {Crottini}, \citenamefont
  {Collin}, \citenamefont {Renucci}, \citenamefont {Pelucchi}, \citenamefont
  {Malko}, \citenamefont {Baier}, \citenamefont {Kapon}, \citenamefont
  {Deveaud},\ and\ \citenamefont {Gani{\`{e}}re}}]{Merano2005}%
  \BibitemOpen
  \bibfield  {author} {\bibinfo {author} {\bibfnamefont {M.}~\bibnamefont
  {Merano}}, \bibinfo {author} {\bibfnamefont {S.}~\bibnamefont {Sonderegger}},
  \bibinfo {author} {\bibfnamefont {A.}~\bibnamefont {Crottini}}, \bibinfo
  {author} {\bibfnamefont {S.}~\bibnamefont {Collin}}, \bibinfo {author}
  {\bibfnamefont {P.}~\bibnamefont {Renucci}}, \bibinfo {author} {\bibfnamefont
  {E.}~\bibnamefont {Pelucchi}}, \bibinfo {author} {\bibfnamefont
  {A.}~\bibnamefont {Malko}}, \bibinfo {author} {\bibfnamefont {M.~H.}\
  \bibnamefont {Baier}}, \bibinfo {author} {\bibfnamefont {E.}~\bibnamefont
  {Kapon}}, \bibinfo {author} {\bibfnamefont {B.}~\bibnamefont {Deveaud}}, \
  and\ \bibinfo {author} {\bibfnamefont {J.-D.}\ \bibnamefont
  {Gani{\`{e}}re}},\ }\bibfield  {title} {\textit {\bibinfo {title} {{Probing
  carrier dynamics in nanostructures by picosecond cathodoluminescence.}}}\
  }\href {\doibase 10.1038/nature04298} {\bibfield  {journal} {\bibinfo
  {journal} {Nature}\ }\textbf {\bibinfo {volume} {438}},\ \bibinfo {pages}
  {479} (\bibinfo {year} {2005})}\BibitemShut {NoStop}%
\bibitem [{\citenamefont {Sonderegger}\ \emph {et~al.}(2006)\citenamefont
  {Sonderegger}, \citenamefont {Feltin}, \citenamefont {Merano}, \citenamefont
  {Crottini}, \citenamefont {Carlin}, \citenamefont {Sachot}, \citenamefont
  {Deveaud}, \citenamefont {Grandjean},\ and\ \citenamefont
  {Gani{\`{e}}re}}]{Sonderegger2006b}%
  \BibitemOpen
  \bibfield  {author} {\bibinfo {author} {\bibfnamefont {S.}~\bibnamefont
  {Sonderegger}}, \bibinfo {author} {\bibfnamefont {E.}~\bibnamefont {Feltin}},
  \bibinfo {author} {\bibfnamefont {M.}~\bibnamefont {Merano}}, \bibinfo
  {author} {\bibfnamefont {A.}~\bibnamefont {Crottini}}, \bibinfo {author}
  {\bibfnamefont {J.~F.}\ \bibnamefont {Carlin}}, \bibinfo {author}
  {\bibfnamefont {R.}~\bibnamefont {Sachot}}, \bibinfo {author} {\bibfnamefont
  {B.}~\bibnamefont {Deveaud}}, \bibinfo {author} {\bibfnamefont
  {N.}~\bibnamefont {Grandjean}}, \ and\ \bibinfo {author} {\bibfnamefont
  {J.~D.}\ \bibnamefont {Gani{\`{e}}re}},\ }\bibfield  {title} {\textit
  {\bibinfo {title} {{High spatial resolution picosecond cathodoluminescence of
  InGaN quantum wells}},}\ }\href {\doibase 10.1063/1.2397562} {\bibfield
  {journal} {\bibinfo  {journal} {Applied Physics Letters}\ }\textbf {\bibinfo
  {volume} {89}},\ \bibinfo {pages} {232109} (\bibinfo {year}
  {2006})}\BibitemShut {NoStop}%
\bibitem [{\citenamefont {Moerland}\ \emph {et~al.}(2016)\citenamefont
  {Moerland}, \citenamefont {Weppelman}, \citenamefont {Garming}, \citenamefont
  {Kruit},\ and\ \citenamefont {Hoogenboom}}]{Moerland2016a}%
  \BibitemOpen
  \bibfield  {author} {\bibinfo {author} {\bibfnamefont {R.~J.}\ \bibnamefont
  {Moerland}}, \bibinfo {author} {\bibfnamefont {I.~G.~C.}\ \bibnamefont
  {Weppelman}}, \bibinfo {author} {\bibfnamefont {M.~W.~H.}\ \bibnamefont
  {Garming}}, \bibinfo {author} {\bibfnamefont {P.}~\bibnamefont {Kruit}}, \
  and\ \bibinfo {author} {\bibfnamefont {J.~P.}\ \bibnamefont {Hoogenboom}},\
  }\bibfield  {title} {\textit {\bibinfo {title} {{Time-resolved
  cathodoluminescence microscopy with sub-nanosecond beam blanking for direct
  evaluation of the local density of states}},}\ }\href {\doibase
  10.1364/OE.24.024760} {\bibfield  {journal} {\bibinfo  {journal} {Optics
  Express}\ }\textbf {\bibinfo {volume} {24}},\ \bibinfo {pages} {24760}
  (\bibinfo {year} {2016})}\BibitemShut {NoStop}%
\bibitem [{\citenamefont {Meuret}\ \emph {et~al.}(2019)\citenamefont {Meuret},
  \citenamefont {{Sol{\`{a}} Garcia}}, \citenamefont {Coenen}, \citenamefont
  {Kieft}, \citenamefont {Zeijlemaker}, \citenamefont {L{\"{a}}tzel},
  \citenamefont {Christiansen}, \citenamefont {Woo}, \citenamefont {Ra},
  \citenamefont {Mi},\ and\ \citenamefont {Polman}}]{Meuret2019}%
  \BibitemOpen
  \bibfield  {author} {\bibinfo {author} {\bibfnamefont {S.}~\bibnamefont
  {Meuret}}, \bibinfo {author} {\bibfnamefont {M.}~\bibnamefont {{Sol{\`{a}}
  Garcia}}}, \bibinfo {author} {\bibfnamefont {T.}~\bibnamefont {Coenen}},
  \bibinfo {author} {\bibfnamefont {E.}~\bibnamefont {Kieft}}, \bibinfo
  {author} {\bibfnamefont {H.}~\bibnamefont {Zeijlemaker}}, \bibinfo {author}
  {\bibfnamefont {M.}~\bibnamefont {L{\"{a}}tzel}}, \bibinfo {author}
  {\bibfnamefont {S.}~\bibnamefont {Christiansen}}, \bibinfo {author}
  {\bibfnamefont {S.}~\bibnamefont {Woo}}, \bibinfo {author} {\bibfnamefont
  {Y.}~\bibnamefont {Ra}}, \bibinfo {author} {\bibfnamefont {Z.}~\bibnamefont
  {Mi}}, \ and\ \bibinfo {author} {\bibfnamefont {A.}~\bibnamefont {Polman}},\
  }\bibfield  {title} {\textit {\bibinfo {title} {{Complementary
  cathodoluminescence lifetime imaging configurations in a scanning electron
  microscope}},}\ }\href {\doibase 10.1016/j.ultramic.2018.11.006} {\bibfield
  {journal} {\bibinfo  {journal} {Ultramicroscopy}\ }\textbf {\bibinfo {volume}
  {197}},\ \bibinfo {pages} {28} (\bibinfo {year} {2019})}\BibitemShut
  {NoStop}%
\bibitem [{\citenamefont {Zewail}(2010)}]{Zewail2010}%
  \BibitemOpen
  \bibfield  {author} {\bibinfo {author} {\bibfnamefont {A.~H.}\ \bibnamefont
  {Zewail}},\ }\bibfield  {title} {\textit {\bibinfo {title} {{Four-Dimensional
  Electron Microscopy}},}\ }\href {\doibase 10.1126/science.1166135} {\bibfield
   {journal} {\bibinfo  {journal} {Science}\ }\textbf {\bibinfo {volume}
  {328}},\ \bibinfo {pages} {187} (\bibinfo {year} {2010})}\BibitemShut
  {NoStop}%
\bibitem [{\citenamefont {Fushitani}(2008)}]{Fushitani2008}%
  \BibitemOpen
  \bibfield  {author} {\bibinfo {author} {\bibfnamefont {M.}~\bibnamefont
  {Fushitani}},\ }\bibfield  {title} {\textit {\bibinfo {title} {{Applications
  of pump-probe spectroscopy}},}\ }\href {\doibase 10.1039/b703983m} {\bibfield
   {journal} {\bibinfo  {journal} {Annual Reports Section "C" (Physical
  Chemistry)}\ }\textbf {\bibinfo {volume} {104}},\ \bibinfo {pages} {272}
  (\bibinfo {year} {2008})}\BibitemShut {NoStop}%
\bibitem [{\citenamefont {Cabanillas-Gonzalez}\ \emph
  {et~al.}(2011)\citenamefont {Cabanillas-Gonzalez}, \citenamefont {Grancini},\
  and\ \citenamefont {Lanzani}}]{Cabanillas-Gonzalez2011}%
  \BibitemOpen
  \bibfield  {author} {\bibinfo {author} {\bibfnamefont {J.}~\bibnamefont
  {Cabanillas-Gonzalez}}, \bibinfo {author} {\bibfnamefont {G.}~\bibnamefont
  {Grancini}}, \ and\ \bibinfo {author} {\bibfnamefont {G.}~\bibnamefont
  {Lanzani}},\ }\bibfield  {title} {\textit {\bibinfo {title} {{Pump-Probe
  Spectroscopy in Organic Semiconductors: Monitoring Fundamental Processes of
  Relevance in Optoelectronics}},}\ }\href {\doibase 10.1002/adma.201102015}
  {\bibfield  {journal} {\bibinfo  {journal} {Advanced Materials}\ }\textbf
  {\bibinfo {volume} {23}},\ \bibinfo {pages} {5468} (\bibinfo {year}
  {2011})}\BibitemShut {NoStop}%
\bibitem [{\citenamefont {Grumstrup}\ \emph {et~al.}(2015)\citenamefont
  {Grumstrup}, \citenamefont {Gabriel}, \citenamefont {Cating}, \citenamefont
  {{Van Goethem}},\ and\ \citenamefont {Papanikolas}}]{Grumstrup2015}%
  \BibitemOpen
  \bibfield  {author} {\bibinfo {author} {\bibfnamefont {E.~M.}\ \bibnamefont
  {Grumstrup}}, \bibinfo {author} {\bibfnamefont {M.~M.}\ \bibnamefont
  {Gabriel}}, \bibinfo {author} {\bibfnamefont {E.~E.}\ \bibnamefont {Cating}},
  \bibinfo {author} {\bibfnamefont {E.~M.}\ \bibnamefont {{Van Goethem}}}, \
  and\ \bibinfo {author} {\bibfnamefont {J.~M.}\ \bibnamefont {Papanikolas}},\
  }\bibfield  {title} {\textit {\bibinfo {title} {{Pump–probe microscopy:
  Visualization and spectroscopy of ultrafast dynamics at the nanoscale}},}\
  }\href {\doibase 10.1016/j.chemphys.2015.07.006} {\bibfield  {journal}
  {\bibinfo  {journal} {Chemical Physics}\ }\textbf {\bibinfo {volume} {458}},\
  \bibinfo {pages} {30} (\bibinfo {year} {2015})}\BibitemShut {NoStop}%
\bibitem [{\citenamefont {Nakajima}\ \emph {et~al.}(2008)\citenamefont
  {Nakajima}, \citenamefont {Takubo}, \citenamefont {Hiroi}, \citenamefont
  {Ueda},\ and\ \citenamefont {Suemoto}}]{Nakajima2008}%
  \BibitemOpen
  \bibfield  {author} {\bibinfo {author} {\bibfnamefont {M.}~\bibnamefont
  {Nakajima}}, \bibinfo {author} {\bibfnamefont {N.}~\bibnamefont {Takubo}},
  \bibinfo {author} {\bibfnamefont {Z.}~\bibnamefont {Hiroi}}, \bibinfo
  {author} {\bibfnamefont {Y.}~\bibnamefont {Ueda}}, \ and\ \bibinfo {author}
  {\bibfnamefont {T.}~\bibnamefont {Suemoto}},\ }\bibfield  {title} {\textit
  {\bibinfo {title} {{Photoinduced metallic state in VO2 proved by the
  terahertz pump-probe spectroscopy}},}\ }\href {\doibase 10.1063/1.2830664}
  {\bibfield  {journal} {\bibinfo  {journal} {Applied Physics Letters}\
  }\textbf {\bibinfo {volume} {92}},\ \bibinfo {pages} {011907} (\bibinfo
  {year} {2008})}\BibitemShut {NoStop}%
\bibitem [{\citenamefont {Pic{\'{o}}n}\ \emph {et~al.}(2016)\citenamefont
  {Pic{\'{o}}n}, \citenamefont {Lehmann}, \citenamefont {Bostedt},
  \citenamefont {Rudenko}, \citenamefont {Marinelli}, \citenamefont {Osipov},
  \citenamefont {Rolles}, \citenamefont {Berrah}, \citenamefont {Bomme},
  \citenamefont {Bucher}, \citenamefont {Doumy}, \citenamefont {Erk},
  \citenamefont {Ferguson}, \citenamefont {Gorkhover}, \citenamefont {Ho},
  \citenamefont {Kanter}, \citenamefont {Kr{\"{a}}ssig}, \citenamefont
  {Krzywinski}, \citenamefont {Lutman}, \citenamefont {March}, \citenamefont
  {Moonshiram}, \citenamefont {Ray}, \citenamefont {Young}, \citenamefont
  {Pratt},\ and\ \citenamefont {Southworth}}]{Picon2016}%
  \BibitemOpen
  \bibfield  {author} {\bibinfo {author} {\bibfnamefont {A.}~\bibnamefont
  {Pic{\'{o}}n}}, \bibinfo {author} {\bibfnamefont {C.~S.}\ \bibnamefont
  {Lehmann}}, \bibinfo {author} {\bibfnamefont {C.}~\bibnamefont {Bostedt}},
  \bibinfo {author} {\bibfnamefont {A.}~\bibnamefont {Rudenko}}, \bibinfo
  {author} {\bibfnamefont {A.}~\bibnamefont {Marinelli}}, \bibinfo {author}
  {\bibfnamefont {T.}~\bibnamefont {Osipov}}, \bibinfo {author} {\bibfnamefont
  {D.}~\bibnamefont {Rolles}}, \bibinfo {author} {\bibfnamefont
  {N.}~\bibnamefont {Berrah}}, \bibinfo {author} {\bibfnamefont
  {C.}~\bibnamefont {Bomme}}, \bibinfo {author} {\bibfnamefont
  {M.}~\bibnamefont {Bucher}}, \bibinfo {author} {\bibfnamefont
  {G.}~\bibnamefont {Doumy}}, \bibinfo {author} {\bibfnamefont
  {B.}~\bibnamefont {Erk}}, \bibinfo {author} {\bibfnamefont {K.~R.}\
  \bibnamefont {Ferguson}}, \bibinfo {author} {\bibfnamefont {T.}~\bibnamefont
  {Gorkhover}}, \bibinfo {author} {\bibfnamefont {P.~J.}\ \bibnamefont {Ho}},
  \bibinfo {author} {\bibfnamefont {E.~P.}\ \bibnamefont {Kanter}}, \bibinfo
  {author} {\bibfnamefont {B.}~\bibnamefont {Kr{\"{a}}ssig}}, \bibinfo {author}
  {\bibfnamefont {J.}~\bibnamefont {Krzywinski}}, \bibinfo {author}
  {\bibfnamefont {A.~A.}\ \bibnamefont {Lutman}}, \bibinfo {author}
  {\bibfnamefont {A.~M.}\ \bibnamefont {March}}, \bibinfo {author}
  {\bibfnamefont {D.}~\bibnamefont {Moonshiram}}, \bibinfo {author}
  {\bibfnamefont {D.}~\bibnamefont {Ray}}, \bibinfo {author} {\bibfnamefont
  {L.}~\bibnamefont {Young}}, \bibinfo {author} {\bibfnamefont {S.~T.}\
  \bibnamefont {Pratt}}, \ and\ \bibinfo {author} {\bibfnamefont {S.~H.}\
  \bibnamefont {Southworth}},\ }\bibfield  {title} {\textit {\bibinfo {title}
  {{Hetero-site-specific X-ray pump-probe spectroscopy for femtosecond
  intramolecular dynamics}},}\ }\href {\doibase 10.1038/ncomms11652} {\bibfield
   {journal} {\bibinfo  {journal} {Nature Communications}\ }\textbf {\bibinfo
  {volume} {7}},\ \bibinfo {pages} {11652} (\bibinfo {year}
  {2016})}\BibitemShut {NoStop}%
\bibitem [{\citenamefont {Murawski}\ \emph {et~al.}(2015)\citenamefont
  {Murawski}, \citenamefont {Graupner}, \citenamefont {Milde}, \citenamefont
  {Raupach}, \citenamefont {Zerweck-Trogisch},\ and\ \citenamefont
  {Eng}}]{Murawski2015}%
  \BibitemOpen
  \bibfield  {author} {\bibinfo {author} {\bibfnamefont {J.}~\bibnamefont
  {Murawski}}, \bibinfo {author} {\bibfnamefont {T.}~\bibnamefont {Graupner}},
  \bibinfo {author} {\bibfnamefont {P.}~\bibnamefont {Milde}}, \bibinfo
  {author} {\bibfnamefont {R.}~\bibnamefont {Raupach}}, \bibinfo {author}
  {\bibfnamefont {U.}~\bibnamefont {Zerweck-Trogisch}}, \ and\ \bibinfo
  {author} {\bibfnamefont {L.~M.}\ \bibnamefont {Eng}},\ }\bibfield  {title}
  {\textit {\bibinfo {title} {{Pump-probe Kelvin-probe force microscopy:
  Principle of operation and resolution limits}},}\ }\href {\doibase
  10.1063/1.4933289} {\bibfield  {journal} {\bibinfo  {journal} {Journal of
  Applied Physics}\ }\textbf {\bibinfo {volume} {118}},\ \bibinfo {pages}
  {154302} (\bibinfo {year} {2015})}\BibitemShut {NoStop}%
\bibitem [{\citenamefont {Jahng}\ \emph {et~al.}(2015)\citenamefont {Jahng},
  \citenamefont {Brocious}, \citenamefont {Fishman}, \citenamefont {Yampolsky},
  \citenamefont {Nowak}, \citenamefont {Huang}, \citenamefont {Apkarian},
  \citenamefont {Wickramasinghe},\ and\ \citenamefont {Potma}}]{Jahng2015}%
  \BibitemOpen
  \bibfield  {author} {\bibinfo {author} {\bibfnamefont {J.}~\bibnamefont
  {Jahng}}, \bibinfo {author} {\bibfnamefont {J.}~\bibnamefont {Brocious}},
  \bibinfo {author} {\bibfnamefont {D.~A.}\ \bibnamefont {Fishman}}, \bibinfo
  {author} {\bibfnamefont {S.}~\bibnamefont {Yampolsky}}, \bibinfo {author}
  {\bibfnamefont {D.}~\bibnamefont {Nowak}}, \bibinfo {author} {\bibfnamefont
  {F.}~\bibnamefont {Huang}}, \bibinfo {author} {\bibfnamefont {V.~A.}\
  \bibnamefont {Apkarian}}, \bibinfo {author} {\bibfnamefont {H.~K.}\
  \bibnamefont {Wickramasinghe}}, \ and\ \bibinfo {author} {\bibfnamefont
  {E.~O.}\ \bibnamefont {Potma}},\ }\bibfield  {title} {\textit {\bibinfo
  {title} {{Ultrafast pump-probe force microscopy with nanoscale
  resolution}},}\ }\href {\doibase 10.1063/1.4913853} {\bibfield  {journal}
  {\bibinfo  {journal} {Applied Physics Letters}\ }\textbf {\bibinfo {volume}
  {106}},\ \bibinfo {pages} {083113} (\bibinfo {year} {2015})}\BibitemShut
  {NoStop}%
\bibitem [{\citenamefont {Lobastov}\ \emph {et~al.}(2005)\citenamefont
  {Lobastov}, \citenamefont {Srinivasan},\ and\ \citenamefont
  {Zewail}}]{Lobastov2005}%
  \BibitemOpen
  \bibfield  {author} {\bibinfo {author} {\bibfnamefont {V.~A.}\ \bibnamefont
  {Lobastov}}, \bibinfo {author} {\bibfnamefont {R.}~\bibnamefont
  {Srinivasan}}, \ and\ \bibinfo {author} {\bibfnamefont {A.~H.}\ \bibnamefont
  {Zewail}},\ }\bibfield  {title} {\textit {\bibinfo {title} {{Four-dimensional
  ultrafast electron microscopy}},}\ }\href {\doibase 10.1073/pnas.0502607102}
  {\bibfield  {journal} {\bibinfo  {journal} {Proceedings of the National
  Academy of Sciences}\ }\textbf {\bibinfo {volume} {102}},\ \bibinfo {pages}
  {7069} (\bibinfo {year} {2005})}\BibitemShut {NoStop}%
\bibitem [{\citenamefont {Barwick}\ \emph {et~al.}(2008)\citenamefont
  {Barwick}, \citenamefont {Park}, \citenamefont {Kwon}, \citenamefont
  {Baskin},\ and\ \citenamefont {Zewail}}]{Barwick2008}%
  \BibitemOpen
  \bibfield  {author} {\bibinfo {author} {\bibfnamefont {B.}~\bibnamefont
  {Barwick}}, \bibinfo {author} {\bibfnamefont {H.~S.}\ \bibnamefont {Park}},
  \bibinfo {author} {\bibfnamefont {O.-h.}\ \bibnamefont {Kwon}}, \bibinfo
  {author} {\bibfnamefont {J.~S.}\ \bibnamefont {Baskin}}, \ and\ \bibinfo
  {author} {\bibfnamefont {A.~H.}\ \bibnamefont {Zewail}},\ }\bibfield  {title}
  {\textit {\bibinfo {title} {{4D Imaging of Transient Structures and
  Morphologies in Ultrafast Electron Microscopy}},}\ }\href {\doibase
  10.1126/science.1164000} {\bibfield  {journal} {\bibinfo  {journal}
  {Science}\ }\textbf {\bibinfo {volume} {322}},\ \bibinfo {pages} {1227}
  (\bibinfo {year} {2008})}\BibitemShut {NoStop}%
\bibitem [{\citenamefont {Mohammed}\ \emph {et~al.}(2011)\citenamefont
  {Mohammed}, \citenamefont {Yang}, \citenamefont {Pal},\ and\ \citenamefont
  {Zewail}}]{Mohammed2011}%
  \BibitemOpen
  \bibfield  {author} {\bibinfo {author} {\bibfnamefont {O.~F.}\ \bibnamefont
  {Mohammed}}, \bibinfo {author} {\bibfnamefont {D.-S.}\ \bibnamefont {Yang}},
  \bibinfo {author} {\bibfnamefont {S.~K.}\ \bibnamefont {Pal}}, \ and\
  \bibinfo {author} {\bibfnamefont {A.~H.}\ \bibnamefont {Zewail}},\ }\bibfield
   {title} {\textit {\bibinfo {title} {{4D Scanning Ultrafast Electron
  Microscopy: Visualization of Materials Surface Dynamics}},}\ }\href {\doibase
  10.1021/ja2031322} {\bibfield  {journal} {\bibinfo  {journal} {Journal of the
  American Chemical Society}\ }\textbf {\bibinfo {volume} {133}},\ \bibinfo
  {pages} {7708} (\bibinfo {year} {2011})}\BibitemShut {NoStop}%
\bibitem [{\citenamefont {Barwick}\ \emph {et~al.}(2009)\citenamefont
  {Barwick}, \citenamefont {Flannigan},\ and\ \citenamefont
  {Zewail}}]{Barwick2009}%
  \BibitemOpen
  \bibfield  {author} {\bibinfo {author} {\bibfnamefont {B.}~\bibnamefont
  {Barwick}}, \bibinfo {author} {\bibfnamefont {D.~J.}\ \bibnamefont
  {Flannigan}}, \ and\ \bibinfo {author} {\bibfnamefont {A.~H.}\ \bibnamefont
  {Zewail}},\ }\bibfield  {title} {\textit {\bibinfo {title} {{Photon-induced
  near-field electron microscopy}},}\ }\href {\doibase 10.1038/nature08662}
  {\bibfield  {journal} {\bibinfo  {journal} {Nature}\ }\textbf {\bibinfo
  {volume} {462}},\ \bibinfo {pages} {902} (\bibinfo {year}
  {2009})}\BibitemShut {NoStop}%
\bibitem [{\citenamefont {{García de Abajo}}\ \emph
  {et~al.}(2010)\citenamefont {{García de Abajo}}, \citenamefont
  {Asenjo-Garcia},\ and\ \citenamefont {Kociak}}]{GarciaDeAbajo2010a}%
  \BibitemOpen
  \bibfield  {author} {\bibinfo {author} {\bibfnamefont {F.~J.}\ \bibnamefont
  {{García de Abajo}}}, \bibinfo {author} {\bibfnamefont {A.}~\bibnamefont
  {Asenjo-Garcia}}, \ and\ \bibinfo {author} {\bibfnamefont {M.}~\bibnamefont
  {Kociak}},\ }\bibfield  {title} {\textit {\bibinfo {title} {{Multiphoton
  Absorption and Emission by Interaction of Swift Electrons with Evanescent
  Light Fields}},}\ }\href {\doibase 10.1021/nl100613s} {\bibfield  {journal}
  {\bibinfo  {journal} {Nano Letters}\ }\textbf {\bibinfo {volume} {10}},\
  \bibinfo {pages} {1859} (\bibinfo {year} {2010})}\BibitemShut {NoStop}%
\bibitem [{\citenamefont {Feist}\ \emph {et~al.}(2015)\citenamefont {Feist},
  \citenamefont {Echternkamp}, \citenamefont {Schauss}, \citenamefont
  {Yalunin}, \citenamefont {Sch{\"{a}}fer},\ and\ \citenamefont
  {Ropers}}]{Feist2015}%
  \BibitemOpen
  \bibfield  {author} {\bibinfo {author} {\bibfnamefont {A.}~\bibnamefont
  {Feist}}, \bibinfo {author} {\bibfnamefont {K.~E.}\ \bibnamefont
  {Echternkamp}}, \bibinfo {author} {\bibfnamefont {J.}~\bibnamefont
  {Schauss}}, \bibinfo {author} {\bibfnamefont {S.~V.}\ \bibnamefont
  {Yalunin}}, \bibinfo {author} {\bibfnamefont {S.}~\bibnamefont
  {Sch{\"{a}}fer}}, \ and\ \bibinfo {author} {\bibfnamefont {C.}~\bibnamefont
  {Ropers}},\ }\bibfield  {title} {\textit {\bibinfo {title} {{Quantum coherent
  optical phase modulation in an ultrafast transmission electron
  microscope}},}\ }\href {\doibase 10.1038/nature14463} {\bibfield  {journal}
  {\bibinfo  {journal} {Nature}\ }\textbf {\bibinfo {volume} {521}},\ \bibinfo
  {pages} {200} (\bibinfo {year} {2015})}\BibitemShut {NoStop}%
\bibitem [{\citenamefont {Carbone}\ \emph {et~al.}(2009)\citenamefont
  {Carbone}, \citenamefont {Kwon},\ and\ \citenamefont
  {Zewail}}]{Carbone2009a}%
  \BibitemOpen
  \bibfield  {author} {\bibinfo {author} {\bibfnamefont {F.}~\bibnamefont
  {Carbone}}, \bibinfo {author} {\bibfnamefont {O.~H.}\ \bibnamefont {Kwon}}, \
  and\ \bibinfo {author} {\bibfnamefont {A.~H.}\ \bibnamefont {Zewail}},\
  }\bibfield  {title} {\textit {\bibinfo {title} {{Dynamics of chemical bonding
  mapped by energy-resolved 4D electron microscopy}},}\ }\href {\doibase
  10.1126/science.1175005} {\bibfield  {journal} {\bibinfo  {journal}
  {Science}\ }\textbf {\bibinfo {volume} {325}},\ \bibinfo {pages} {181}
  (\bibinfo {year} {2009})}\BibitemShut {NoStop}%
\bibitem [{\citenamefont {Schliep}\ \emph {et~al.}(2017)\citenamefont
  {Schliep}, \citenamefont {Quarterman}, \citenamefont {Wang},\ and\
  \citenamefont {Flannigan}}]{Schliep2017}%
  \BibitemOpen
  \bibfield  {author} {\bibinfo {author} {\bibfnamefont {K.~B.}\ \bibnamefont
  {Schliep}}, \bibinfo {author} {\bibfnamefont {P.}~\bibnamefont {Quarterman}},
  \bibinfo {author} {\bibfnamefont {J.-P.}\ \bibnamefont {Wang}}, \ and\
  \bibinfo {author} {\bibfnamefont {D.~J.}\ \bibnamefont {Flannigan}},\
  }\bibfield  {title} {\textit {\bibinfo {title} {{Picosecond Fresnel
  transmission electron microscopy}},}\ }\href {\doibase 10.1063/1.4984586}
  {\bibfield  {journal} {\bibinfo  {journal} {Applied Physics Letters}\
  }\textbf {\bibinfo {volume} {110}},\ \bibinfo {pages} {222404} (\bibinfo
  {year} {2017})}\BibitemShut {NoStop}%
\bibitem [{\citenamefont {{Rubiano da Silva}}\ \emph
  {et~al.}(2018)\citenamefont {{Rubiano da Silva}}, \citenamefont
  {M{\"{o}}ller}, \citenamefont {Feist}, \citenamefont {Ulrichs}, \citenamefont
  {Ropers},\ and\ \citenamefont {Sch{\"{a}}fer}}]{RubianoDaSilva2018}%
  \BibitemOpen
  \bibfield  {author} {\bibinfo {author} {\bibfnamefont {N.}~\bibnamefont
  {{Rubiano da Silva}}}, \bibinfo {author} {\bibfnamefont {M.}~\bibnamefont
  {M{\"{o}}ller}}, \bibinfo {author} {\bibfnamefont {A.}~\bibnamefont {Feist}},
  \bibinfo {author} {\bibfnamefont {H.}~\bibnamefont {Ulrichs}}, \bibinfo
  {author} {\bibfnamefont {C.}~\bibnamefont {Ropers}}, \ and\ \bibinfo {author}
  {\bibfnamefont {S.}~\bibnamefont {Sch{\"{a}}fer}},\ }\bibfield  {title}
  {\textit {\bibinfo {title} {{Nanoscale Mapping of Ultrafast Magnetization
  Dynamics with Femtosecond Lorentz Microscopy}},}\ }\href {\doibase
  10.1103/PhysRevX.8.031052} {\bibfield  {journal} {\bibinfo  {journal}
  {Physical Review X}\ }\textbf {\bibinfo {volume} {8}},\ \bibinfo {pages}
  {031052} (\bibinfo {year} {2018})}\BibitemShut {NoStop}%
\bibitem [{\citenamefont {Cremons}\ \emph {et~al.}(2016)\citenamefont
  {Cremons}, \citenamefont {Plemmons},\ and\ \citenamefont
  {Flannigan}}]{Cremons2016}%
  \BibitemOpen
  \bibfield  {author} {\bibinfo {author} {\bibfnamefont {D.~R.}\ \bibnamefont
  {Cremons}}, \bibinfo {author} {\bibfnamefont {D.~A.}\ \bibnamefont
  {Plemmons}}, \ and\ \bibinfo {author} {\bibfnamefont {D.~J.}\ \bibnamefont
  {Flannigan}},\ }\bibfield  {title} {\textit {\bibinfo {title} {{Femtosecond
  electron imaging of defect-modulated phonon dynamics}},}\ }\href {\doibase
  10.1038/ncomms11230} {\bibfield  {journal} {\bibinfo  {journal} {Nature
  Communications}\ }\textbf {\bibinfo {volume} {7}},\ \bibinfo {pages} {11230}
  (\bibinfo {year} {2016})}\BibitemShut {NoStop}%
\bibitem [{\citenamefont {Valley}\ \emph {et~al.}(2016)\citenamefont {Valley},
  \citenamefont {Ferry},\ and\ \citenamefont {Flannigan}}]{Valley2016}%
  \BibitemOpen
  \bibfield  {author} {\bibinfo {author} {\bibfnamefont {D.~T.}\ \bibnamefont
  {Valley}}, \bibinfo {author} {\bibfnamefont {V.~E.}\ \bibnamefont {Ferry}}, \
  and\ \bibinfo {author} {\bibfnamefont {D.~J.}\ \bibnamefont {Flannigan}},\
  }\bibfield  {title} {\textit {\bibinfo {title} {{Imaging Intra- and
  Interparticle Acousto-plasmonic Vibrational Dynamics with Ultrafast Electron
  Microscopy}},}\ }\href {\doibase 10.1021/acs.nanolett.6b03975} {\bibfield
  {journal} {\bibinfo  {journal} {Nano Letters}\ }\textbf {\bibinfo {volume}
  {16}},\ \bibinfo {pages} {7302} (\bibinfo {year} {2016})}\BibitemShut
  {NoStop}%
\bibitem [{\citenamefont {Liao}\ and\ \citenamefont
  {Najafi}(2017)}]{Liao2017a}%
  \BibitemOpen
  \bibfield  {author} {\bibinfo {author} {\bibfnamefont {B.}~\bibnamefont
  {Liao}}\ and\ \bibinfo {author} {\bibfnamefont {E.}~\bibnamefont {Najafi}},\
  }\bibfield  {title} {\textit {\bibinfo {title} {{Scanning ultrafast electron
  microscopy: A novel technique to probe photocarrier dynamics with high
  spatial and temporal resolutions}},}\ }\href {\doibase
  10.1016/j.mtphys.2017.07.003} {\bibfield  {journal} {\bibinfo  {journal}
  {Materials Today Physics}\ }\textbf {\bibinfo {volume} {2}},\ \bibinfo
  {pages} {46} (\bibinfo {year} {2017})}\BibitemShut {NoStop}%
\bibitem [{\citenamefont {Garming}\ \emph {et~al.}(2020)\citenamefont
  {Garming}, \citenamefont {Bolhuis}, \citenamefont {Conesa-Boj}, \citenamefont
  {Kruit},\ and\ \citenamefont {Hoogenboom}}]{Garming2020}%
  \BibitemOpen
  \bibfield  {author} {\bibinfo {author} {\bibfnamefont {M.~W.~H.}\
  \bibnamefont {Garming}}, \bibinfo {author} {\bibfnamefont {M.}~\bibnamefont
  {Bolhuis}}, \bibinfo {author} {\bibfnamefont {S.}~\bibnamefont {Conesa-Boj}},
  \bibinfo {author} {\bibfnamefont {P.}~\bibnamefont {Kruit}}, \ and\ \bibinfo
  {author} {\bibfnamefont {J.~P.}\ \bibnamefont {Hoogenboom}},\ }\bibfield
  {title} {\textit {\bibinfo {title} {{Lock-in Ultrafast Electron Microscopy
  Simultaneously Visualizes Carrier Recombination and Interface-Mediated
  Trapping}},}\ }\href {\doibase 10.1021/acs.jpclett.0c02345} {\bibfield
  {journal} {\bibinfo  {journal} {The Journal of Physical Chemistry Letters}\
  }\textbf {\bibinfo {volume} {11}},\ \bibinfo {pages} {8880} (\bibinfo {year}
  {2020})}\BibitemShut {NoStop}%
\bibitem [{\citenamefont {Sol{\`{a}}-Garcia}\ \emph {et~al.}(2020)\citenamefont
  {Sol{\`{a}}-Garcia}, \citenamefont {Meuret}, \citenamefont {Coenen},\ and\
  \citenamefont {Polman}}]{Sola-Garcia2019}%
  \BibitemOpen
  \bibfield  {author} {\bibinfo {author} {\bibfnamefont {M.}~\bibnamefont
  {Sol{\`{a}}-Garcia}}, \bibinfo {author} {\bibfnamefont {S.}~\bibnamefont
  {Meuret}}, \bibinfo {author} {\bibfnamefont {T.}~\bibnamefont {Coenen}}, \
  and\ \bibinfo {author} {\bibfnamefont {A.}~\bibnamefont {Polman}},\
  }\bibfield  {title} {\textit {\bibinfo {title} {{Electron-Induced State
  Conversion in Diamond NV Centers Measured with Pump–Probe
  Cathodoluminescence Spectroscopy}},}\ }\href {\doibase
  10.1021/acsphotonics.9b01463} {\bibfield  {journal} {\bibinfo  {journal} {ACS
  Photonics}\ }\textbf {\bibinfo {volume} {7}},\ \bibinfo {pages} {232}
  (\bibinfo {year} {2020})}\BibitemShut {NoStop}%
\bibitem [{\citenamefont {Arbouet}\ \emph {et~al.}(2018)\citenamefont
  {Arbouet}, \citenamefont {Caruso},\ and\ \citenamefont
  {Houdellier}}]{Arbouet2018}%
  \BibitemOpen
  \bibfield  {author} {\bibinfo {author} {\bibfnamefont {A.}~\bibnamefont
  {Arbouet}}, \bibinfo {author} {\bibfnamefont {G.~M.}\ \bibnamefont {Caruso}},
  \ and\ \bibinfo {author} {\bibfnamefont {F.}~\bibnamefont {Houdellier}},\
  }\bibfield  {title} {\textit {\bibinfo {title} {{Ultrafast Transmission
  Electron Microscopy: Historical Development, Instrumentation, and
  Applications}},}\ }in\ \href {\doibase 10.1016/bs.aiep.2018.06.001} {\emph
  {\bibinfo {booktitle} {Advances in Imaging and Electron Physics, Volume
  207}}}\ (\bibinfo  {publisher} {Elsevier},\ \bibinfo {year} {2018})\
  Chap.~\bibinfo {chapter} {1}, pp.\ \bibinfo {pages} {1--72}\BibitemShut
  {NoStop}%
\bibitem [{\citenamefont {Coenen}\ \emph {et~al.}(2011)\citenamefont {Coenen},
  \citenamefont {Vesseur},\ and\ \citenamefont {Polman}}]{Coenen2011}%
  \BibitemOpen
  \bibfield  {author} {\bibinfo {author} {\bibfnamefont {T.}~\bibnamefont
  {Coenen}}, \bibinfo {author} {\bibfnamefont {E.~J.~R.}\ \bibnamefont
  {Vesseur}}, \ and\ \bibinfo {author} {\bibfnamefont {A.}~\bibnamefont
  {Polman}},\ }\bibfield  {title} {\textit {\bibinfo {title} {{Angle-resolved
  cathodoluminescence spectroscopy}},}\ }\href {\doibase 10.1063/1.3644985}
  {\bibfield  {journal} {\bibinfo  {journal} {Applied Physics Letters}\
  }\textbf {\bibinfo {volume} {99}},\ \bibinfo {pages} {143103} (\bibinfo
  {year} {2011})}\BibitemShut {NoStop}%
\bibitem [{\citenamefont {Feist}\ \emph {et~al.}(2017)\citenamefont {Feist},
  \citenamefont {Bach}, \citenamefont {{Rubiano da Silva}}, \citenamefont
  {Danz}, \citenamefont {M{\"{o}}ller}, \citenamefont {Priebe}, \citenamefont
  {Domr{\"{o}}se}, \citenamefont {Gatzmann}, \citenamefont {Rost},
  \citenamefont {Schauss}, \citenamefont {Strauch}, \citenamefont {Bormann},
  \citenamefont {Sivis}, \citenamefont {Sch{\"{a}}fer},\ and\ \citenamefont
  {Ropers}}]{Feist2017a}%
  \BibitemOpen
  \bibfield  {author} {\bibinfo {author} {\bibfnamefont {A.}~\bibnamefont
  {Feist}}, \bibinfo {author} {\bibfnamefont {N.}~\bibnamefont {Bach}},
  \bibinfo {author} {\bibfnamefont {N.}~\bibnamefont {{Rubiano da Silva}}},
  \bibinfo {author} {\bibfnamefont {T.}~\bibnamefont {Danz}}, \bibinfo {author}
  {\bibfnamefont {M.}~\bibnamefont {M{\"{o}}ller}}, \bibinfo {author}
  {\bibfnamefont {K.~E.}\ \bibnamefont {Priebe}}, \bibinfo {author}
  {\bibfnamefont {T.}~\bibnamefont {Domr{\"{o}}se}}, \bibinfo {author}
  {\bibfnamefont {J.~G.}\ \bibnamefont {Gatzmann}}, \bibinfo {author}
  {\bibfnamefont {S.}~\bibnamefont {Rost}}, \bibinfo {author} {\bibfnamefont
  {J.}~\bibnamefont {Schauss}}, \bibinfo {author} {\bibfnamefont
  {S.}~\bibnamefont {Strauch}}, \bibinfo {author} {\bibfnamefont
  {R.}~\bibnamefont {Bormann}}, \bibinfo {author} {\bibfnamefont
  {M.}~\bibnamefont {Sivis}}, \bibinfo {author} {\bibfnamefont
  {S.}~\bibnamefont {Sch{\"{a}}fer}}, \ and\ \bibinfo {author} {\bibfnamefont
  {C.}~\bibnamefont {Ropers}},\ }\bibfield  {title} {\textit {\bibinfo {title}
  {{Ultrafast transmission electron microscopy using a laser-driven field
  emitter: Femtosecond resolution with a high coherence electron beam}},}\
  }\href {\doibase 10.1016/j.ultramic.2016.12.005} {\bibfield  {journal}
  {\bibinfo  {journal} {Ultramicroscopy}\ }\textbf {\bibinfo {volume} {176}},\
  \bibinfo {pages} {63} (\bibinfo {year} {2017})}\BibitemShut {NoStop}%
\bibitem [{\citenamefont {Sol{\`{a}}-Garcia}(2021)}]{Sola-Garcia2021a}%
  \BibitemOpen
  \bibfield  {author} {\bibinfo {author} {\bibfnamefont {M.}~\bibnamefont
  {Sol{\`{a}}-Garcia}},\ }\emph {\bibinfo {title} {{Electron-matter interaction
  probed with time-resolved cathodoluminescence}}},\ \href
  {https://ir.amolf.nl/pub/10407} {Ph.D. thesis},\ \bibinfo  {school}
  {University of Amsterdam} (\bibinfo {year} {2021})\BibitemShut {NoStop}%
\bibitem [{\citenamefont {Bronsgeest}(2009)}]{Bronsgeest2009}%
  \BibitemOpen
  \bibfield  {author} {\bibinfo {author} {\bibfnamefont {M.}~\bibnamefont
  {Bronsgeest}},\ }\emph {\bibinfo {title} {{Physics of Schottky electron
  sources}}},\ \href
  {https://repository.tudelft.nl/islandora/object/uuid:7975ef5e-c2ea-4056-be43-6bb6c062c884?collection=research}
  {Ph.D. thesis},\ \bibinfo  {school} {Delft University of Technology}
  (\bibinfo {year} {2009})\BibitemShut {NoStop}%
\bibitem [{\citenamefont {Coenen}(2014)}]{Coenen2014}%
  \BibitemOpen
  \bibfield  {author} {\bibinfo {author} {\bibfnamefont {T.}~\bibnamefont
  {Coenen}},\ }\emph {\bibinfo {title} {{Angle-resolved cathodoluminescence
  nanoscopy}}},\ \href {https://hdl.handle.net/11245/1.417564} {Ph.D. thesis},\
  \bibinfo  {school} {University of Amsterdam} (\bibinfo {year}
  {2014})\BibitemShut {NoStop}%
\bibitem [{\citenamefont {Brenny}(2016)}]{Brenny2016a}%
  \BibitemOpen
  \bibfield  {author} {\bibinfo {author} {\bibfnamefont {B.~J.}\ \bibnamefont
  {Brenny}},\ }\emph {\bibinfo {title} {{Probing light emission at the
  nanoscale with cathodoluminescence}}},\ \href@noop {} {Ph.D. thesis},\
  \bibinfo  {school} {University of Amsterdam} (\bibinfo {year}
  {2016})\BibitemShut {NoStop}%
\bibitem [{\citenamefont {Vesseur}(2011)}]{Vesseur2011}%
  \BibitemOpen
  \bibfield  {author} {\bibinfo {author} {\bibfnamefont {E.~J.~R.}\
  \bibnamefont {Vesseur}},\ }\emph {\bibinfo {title} {{Electron beam imaging
  and spectroscopy of plasmonic nanoantenna resonances}}},\ \href@noop {}
  {Ph.D. thesis},\ \bibinfo  {school} {Utrecht University} (\bibinfo {year}
  {2011})\BibitemShut {NoStop}%
\bibitem [{\citenamefont {Brenny}\ \emph {et~al.}(2014)\citenamefont {Brenny},
  \citenamefont {Coenen},\ and\ \citenamefont {Polman}}]{Brenny2014}%
  \BibitemOpen
  \bibfield  {author} {\bibinfo {author} {\bibfnamefont {B.~J.~M.}\
  \bibnamefont {Brenny}}, \bibinfo {author} {\bibfnamefont {T.}~\bibnamefont
  {Coenen}}, \ and\ \bibinfo {author} {\bibfnamefont {A.}~\bibnamefont
  {Polman}},\ }\bibfield  {title} {\textit {\bibinfo {title} {{Quantifying
  coherent and incoherent cathodoluminescence in semiconductors and metals}},}\
  }\href {\doibase 10.1063/1.4885426} {\bibfield  {journal} {\bibinfo
  {journal} {Journal of Applied Physics}\ }\textbf {\bibinfo {volume} {115}},\
  \bibinfo {pages} {244307} (\bibinfo {year} {2014})}\BibitemShut {NoStop}%
\bibitem [{\citenamefont {Meuret}\ \emph {et~al.}(2017)\citenamefont {Meuret},
  \citenamefont {Coenen}, \citenamefont {Zeijlemaker}, \citenamefont {Latzel},
  \citenamefont {Christiansen}, \citenamefont {Conesa-Boj},\ and\ \citenamefont
  {Polman}}]{Meuret2017}%
  \BibitemOpen
  \bibfield  {author} {\bibinfo {author} {\bibfnamefont {S.}~\bibnamefont
  {Meuret}}, \bibinfo {author} {\bibfnamefont {T.}~\bibnamefont {Coenen}},
  \bibinfo {author} {\bibfnamefont {H.}~\bibnamefont {Zeijlemaker}}, \bibinfo
  {author} {\bibfnamefont {M.}~\bibnamefont {Latzel}}, \bibinfo {author}
  {\bibfnamefont {S.}~\bibnamefont {Christiansen}}, \bibinfo {author}
  {\bibfnamefont {S.}~\bibnamefont {Conesa-Boj}}, \ and\ \bibinfo {author}
  {\bibfnamefont {A.}~\bibnamefont {Polman}},\ }\bibfield  {title} {\textit
  {\bibinfo {title} {{Photon bunching reveals single-electron
  cathodoluminescence excitation efficiency in InGaN quantum wells}},}\ }\href
  {\doibase 10.1103/PhysRevB.96.035308} {\bibfield  {journal} {\bibinfo
  {journal} {Physical Review B}\ }\textbf {\bibinfo {volume} {96}},\ \bibinfo
  {pages} {035308} (\bibinfo {year} {2017})}\BibitemShut {NoStop}%
\bibitem [{\citenamefont {{Garc{\'{i}}a de Abajo}}(2010)}]{GarciaDeAbajo2010}%
  \BibitemOpen
  \bibfield  {author} {\bibinfo {author} {\bibfnamefont {F.~J.}\ \bibnamefont
  {{Garc{\'{i}}a de Abajo}}},\ }\bibfield  {title} {\textit {\bibinfo {title}
  {{Optical excitations in electron microscopy}},}\ }\href {\doibase
  10.1103/RevModPhys.82.209} {\bibfield  {journal} {\bibinfo  {journal}
  {Reviews of Modern Physics}\ }\textbf {\bibinfo {volume} {82}},\ \bibinfo
  {pages} {209} (\bibinfo {year} {2010})}\BibitemShut {NoStop}%
\bibitem [{\citenamefont {Wahl}(2014)}]{Wahl2017}%
  \BibitemOpen
  \bibfield  {author} {\bibinfo {author} {\bibfnamefont {M.}~\bibnamefont
  {Wahl}},\ }\href@noop {} {\emph {\bibinfo {title} {{Time-Correlated Single
  Photon Counting}}}},\ \bibinfo {type} {Tech. Rep.}\ (\bibinfo  {institution}
  {PicoQuant},\ \bibinfo {year} {2014})\BibitemShut {NoStop}%
\bibitem [{\citenamefont {Brenny}\ \emph {et~al.}(2016)\citenamefont {Brenny},
  \citenamefont {Polman},\ and\ \citenamefont {{Garc{\'{i}}a De
  Abajo}}}]{Brenny2016}%
  \BibitemOpen
  \bibfield  {author} {\bibinfo {author} {\bibfnamefont {B.~J.}\ \bibnamefont
  {Brenny}}, \bibinfo {author} {\bibfnamefont {A.}~\bibnamefont {Polman}}, \
  and\ \bibinfo {author} {\bibfnamefont {F.~J.}\ \bibnamefont {{Garc{\'{i}}a De
  Abajo}}},\ }\bibfield  {title} {\textit {\bibinfo {title} {{Femtosecond
  plasmon and photon wave packets excited by a high-energy electron on a metal
  or dielectric surface}},}\ }\href {\doibase 10.1103/PhysRevB.94.155412}
  {\bibfield  {journal} {\bibinfo  {journal} {Physical Review B}\ }\textbf
  {\bibinfo {volume} {94}},\ \bibinfo {pages} {1} (\bibinfo {year}
  {2016})}\BibitemShut {NoStop}%
\bibitem [{\citenamefont {Sun}\ \emph {et~al.}(2016)\citenamefont {Sun},
  \citenamefont {Adhikari}, \citenamefont {Shaheen}, \citenamefont {Yang},\
  and\ \citenamefont {Mohammed}}]{Sun2016a}%
  \BibitemOpen
  \bibfield  {author} {\bibinfo {author} {\bibfnamefont {J.}~\bibnamefont
  {Sun}}, \bibinfo {author} {\bibfnamefont {A.}~\bibnamefont {Adhikari}},
  \bibinfo {author} {\bibfnamefont {B.~S.}\ \bibnamefont {Shaheen}}, \bibinfo
  {author} {\bibfnamefont {H.}~\bibnamefont {Yang}}, \ and\ \bibinfo {author}
  {\bibfnamefont {O.~F.}\ \bibnamefont {Mohammed}},\ }\bibfield  {title}
  {\textit {\bibinfo {title} {{Mapping Carrier Dynamics on Material Surfaces in
  Space and Time using Scanning Ultrafast Electron Microscopy}},}\ }\href
  {\doibase 10.1021/acs.jpclett.5b02908} {\bibfield  {journal} {\bibinfo
  {journal} {The Journal of Physical Chemistry Letters}\ }\textbf {\bibinfo
  {volume} {7}},\ \bibinfo {pages} {985} (\bibinfo {year} {2016})}\BibitemShut
  {NoStop}%
\bibitem [{\citenamefont {Fox}(2006)}]{Fox2006}%
  \BibitemOpen
  \bibfield  {author} {\bibinfo {author} {\bibfnamefont {M.}~\bibnamefont
  {Fox}},\ }\href@noop {} {\emph {\bibinfo {title} {{Quantum Optics: An
  Introduction}}}}\ (\bibinfo  {publisher} {Oxford University Press},\ \bibinfo
  {year} {2006})\ p.\ \bibinfo {pages} {397}\BibitemShut {NoStop}%
\bibitem [{\citenamefont {{Hanbury Brown}}\ and\ \citenamefont
  {Twiss}(1956)}]{HanburyBrown1956a}%
  \BibitemOpen
  \bibfield  {author} {\bibinfo {author} {\bibfnamefont {R.}~\bibnamefont
  {{Hanbury Brown}}}\ and\ \bibinfo {author} {\bibfnamefont {R.~Q.}\
  \bibnamefont {Twiss}},\ }\bibfield  {title} {\textit {\bibinfo {title}
  {{Correlation between photons in two coherent beams of light}},}\ }\href
  {\doibase 10.1007/BF03010401} {\bibfield  {journal} {\bibinfo  {journal}
  {Nature}\ }\textbf {\bibinfo {volume} {177}},\ \bibinfo {pages} {27}
  (\bibinfo {year} {1956})}\BibitemShut {NoStop}%
\bibitem [{\citenamefont {Sol{\`{a}}-Garcia}\ \emph {et~al.}(2021)\citenamefont
  {Sol{\`{a}}-Garcia}, \citenamefont {Mauser}, \citenamefont {Liebtrau},
  \citenamefont {Coenen}, \citenamefont {Christiansen}, \citenamefont
  {Meuret},\ and\ \citenamefont {Polman}}]{Sola-Garcia2021}%
  \BibitemOpen
  \bibfield  {author} {\bibinfo {author} {\bibfnamefont {M.}~\bibnamefont
  {Sol{\`{a}}-Garcia}}, \bibinfo {author} {\bibfnamefont {K.~W.}\ \bibnamefont
  {Mauser}}, \bibinfo {author} {\bibfnamefont {M.}~\bibnamefont {Liebtrau}},
  \bibinfo {author} {\bibfnamefont {T.}~\bibnamefont {Coenen}}, \bibinfo
  {author} {\bibfnamefont {S.}~\bibnamefont {Christiansen}}, \bibinfo {author}
  {\bibfnamefont {S.}~\bibnamefont {Meuret}}, \ and\ \bibinfo {author}
  {\bibfnamefont {A.}~\bibnamefont {Polman}},\ }\bibfield  {title} {\textit
  {\bibinfo {title} {{Photon Statistics of Incoherent Cathodoluminescence with
  Continuous and Pulsed Electron Beams}},}\ }\href {\doibase
  10.1021/acsphotonics.0c01939} {\bibfield  {journal} {\bibinfo  {journal} {ACS
  Photonics}\ }\textbf {\bibinfo {volume} {8}},\ \bibinfo {pages} {916}
  (\bibinfo {year} {2021})},\ \Eprint {http://arxiv.org/abs/2012.12375}
  {arXiv:2012.12375} \BibitemShut {NoStop}%
\bibitem [{\citenamefont {Rech}\ \emph {et~al.}(2008)\citenamefont {Rech},
  \citenamefont {Ingargiola}, \citenamefont {Spinelli}, \citenamefont
  {Labanca}, \citenamefont {Marangoni}, \citenamefont {Ghioni},\ and\
  \citenamefont {Cova}}]{Rech2008}%
  \BibitemOpen
  \bibfield  {author} {\bibinfo {author} {\bibfnamefont {I.}~\bibnamefont
  {Rech}}, \bibinfo {author} {\bibfnamefont {A.}~\bibnamefont {Ingargiola}},
  \bibinfo {author} {\bibfnamefont {R.}~\bibnamefont {Spinelli}}, \bibinfo
  {author} {\bibfnamefont {I.}~\bibnamefont {Labanca}}, \bibinfo {author}
  {\bibfnamefont {S.}~\bibnamefont {Marangoni}}, \bibinfo {author}
  {\bibfnamefont {M.}~\bibnamefont {Ghioni}}, \ and\ \bibinfo {author}
  {\bibfnamefont {S.}~\bibnamefont {Cova}},\ }\bibfield  {title} {\textit
  {\bibinfo {title} {{Optical crosstalk in single photon avalanche diode
  arrays: a new complete model}},}\ }\href {\doibase 10.1364/OE.16.008381}
  {\bibfield  {journal} {\bibinfo  {journal} {Optics Express}\ }\textbf
  {\bibinfo {volume} {16}},\ \bibinfo {pages} {8381} (\bibinfo {year}
  {2008})}\BibitemShut {NoStop}%
\bibitem [{\citenamefont {Howard}(1979)}]{Howard1979}%
  \BibitemOpen
  \bibfield  {author} {\bibinfo {author} {\bibfnamefont {J.~E.}\ \bibnamefont
  {Howard}},\ }\bibfield  {title} {\textit {\bibinfo {title} {{Imaging
  properties of off-axis parabolic mirrors}},}\ }\href {\doibase
  10.1364/ao.18.002714} {\bibfield  {journal} {\bibinfo  {journal} {Applied
  Optics}\ }\textbf {\bibinfo {volume} {18}},\ \bibinfo {pages} {2714}
  (\bibinfo {year} {1979})}\BibitemShut {NoStop}%
\bibitem [{\citenamefont {Drechsler}\ \emph {et~al.}(2001)\citenamefont
  {Drechsler}, \citenamefont {Lieb}, \citenamefont {Debus}, \citenamefont
  {Meixner},\ and\ \citenamefont {Tarrach}}]{Drechsler2001}%
  \BibitemOpen
  \bibfield  {author} {\bibinfo {author} {\bibfnamefont {A.}~\bibnamefont
  {Drechsler}}, \bibinfo {author} {\bibfnamefont {M.}~\bibnamefont {Lieb}},
  \bibinfo {author} {\bibfnamefont {C.}~\bibnamefont {Debus}}, \bibinfo
  {author} {\bibfnamefont {A.}~\bibnamefont {Meixner}}, \ and\ \bibinfo
  {author} {\bibfnamefont {G.}~\bibnamefont {Tarrach}},\ }\bibfield  {title}
  {\textit {\bibinfo {title} {{Confocal microscopy with a high numerical
  aperture parabolic mirror}},}\ }\href {\doibase 10.1364/OE.9.000637}
  {\bibfield  {journal} {\bibinfo  {journal} {Optics Express}\ }\textbf
  {\bibinfo {volume} {9}},\ \bibinfo {pages} {637} (\bibinfo {year}
  {2001})}\BibitemShut {NoStop}%
\bibitem [{\citenamefont {Eliseev}\ \emph {et~al.}(1999)\citenamefont
  {Eliseev}, \citenamefont {Sun}, \citenamefont {Juodkazis}, \citenamefont
  {Sugahara}, \citenamefont {Sakai},\ and\ \citenamefont
  {Misawa}}]{Eliseev1999}%
  \BibitemOpen
  \bibfield  {author} {\bibinfo {author} {\bibfnamefont {P.~G.}\ \bibnamefont
  {Eliseev}}, \bibinfo {author} {\bibfnamefont {H.-B.}\ \bibnamefont {Sun}},
  \bibinfo {author} {\bibfnamefont {S.}~\bibnamefont {Juodkazis}}, \bibinfo
  {author} {\bibfnamefont {T.}~\bibnamefont {Sugahara}}, \bibinfo {author}
  {\bibfnamefont {S.}~\bibnamefont {Sakai}}, \ and\ \bibinfo {author}
  {\bibfnamefont {H.}~\bibnamefont {Misawa}},\ }\bibfield  {title} {\textit
  {\bibinfo {title} {{Laser-Induced Damage Threshold and Surface Processing of
  GaN at 400 nm Wavelength}},}\ }\href {\doibase 10.1143/JJAP.38.L839}
  {\bibfield  {journal} {\bibinfo  {journal} {Japanese Journal of Applied
  Physics}\ }\textbf {\bibinfo {volume} {38}},\ \bibinfo {pages} {L839}
  (\bibinfo {year} {1999})}\BibitemShut {NoStop}%
\bibitem [{\citenamefont {Akane}\ \emph {et~al.}(1999)\citenamefont {Akane},
  \citenamefont {Sugioka}, \citenamefont {Ogino}, \citenamefont {Takai},\ and\
  \citenamefont {Midorikawa}}]{Akane1999}%
  \BibitemOpen
  \bibfield  {author} {\bibinfo {author} {\bibfnamefont {T.}~\bibnamefont
  {Akane}}, \bibinfo {author} {\bibfnamefont {K.}~\bibnamefont {Sugioka}},
  \bibinfo {author} {\bibfnamefont {H.}~\bibnamefont {Ogino}}, \bibinfo
  {author} {\bibfnamefont {H.}~\bibnamefont {Takai}}, \ and\ \bibinfo {author}
  {\bibfnamefont {K.}~\bibnamefont {Midorikawa}},\ }\bibfield  {title} {\textit
  {\bibinfo {title} {{KrF excimer laser induced ablation–planarization of GaN
  surface}},}\ }\href {\doibase 10.1016/S0169-4332(99)00156-7} {\bibfield
  {journal} {\bibinfo  {journal} {Applied Surface Science}\ }\textbf {\bibinfo
  {volume} {148}},\ \bibinfo {pages} {133} (\bibinfo {year}
  {1999})}\BibitemShut {NoStop}%
\bibitem [{\citenamefont {Dubowski}\ \emph {et~al.}(2001)\citenamefont
  {Dubowski}, \citenamefont {Moisa}, \citenamefont {Komorowski}, \citenamefont
  {Tang},\ and\ \citenamefont {Webb}}]{Dubowski2001}%
  \BibitemOpen
  \bibfield  {author} {\bibinfo {author} {\bibfnamefont {J.~J.}\ \bibnamefont
  {Dubowski}}, \bibinfo {author} {\bibfnamefont {S.}~\bibnamefont {Moisa}},
  \bibinfo {author} {\bibfnamefont {B.}~\bibnamefont {Komorowski}}, \bibinfo
  {author} {\bibfnamefont {H.}~\bibnamefont {Tang}}, \ and\ \bibinfo {author}
  {\bibfnamefont {J.~B.}\ \bibnamefont {Webb}},\ }\bibfield  {title} {\textit
  {\bibinfo {title} {{Laser polishing of GaN}},}\ }in\ \href {\doibase
  10.1117/12.432536} {\emph {\bibinfo {booktitle} {Proc. SPIE 4274, Laser
  Applications in Microelectronic and Optoelectronic Manufacturing VI,}}},\
  \bibinfo {editor} {edited by\ \bibinfo {editor} {\bibfnamefont {M.~C.}\
  \bibnamefont {Gower}}, \bibinfo {editor} {\bibfnamefont {H.}~\bibnamefont
  {Helvajian}}, \bibinfo {editor} {\bibfnamefont {K.}~\bibnamefont {Sugioka}},
  \ and\ \bibinfo {editor} {\bibfnamefont {J.~J.}\ \bibnamefont {Dubowski}}}\
  (\bibinfo {year} {2001})\ pp.\ \bibinfo {pages} {442--447}\BibitemShut
  {NoStop}%
\bibitem [{\citenamefont {Vladar}\ and\ \citenamefont
  {Postek}(2005)}]{Vladar2005}%
  \BibitemOpen
  \bibfield  {author} {\bibinfo {author} {\bibfnamefont {A.}~\bibnamefont
  {Vladar}}\ and\ \bibinfo {author} {\bibfnamefont {M.}~\bibnamefont
  {Postek}},\ }\bibfield  {title} {\textit {\bibinfo {title} {{Electron
  Beam-Induced Sample Contamination in the SEM}},}\ }\href {\doibase
  10.1017/S1431927605507785} {\bibfield  {journal} {\bibinfo  {journal}
  {Microscopy and Microanalysis}\ }\textbf {\bibinfo {volume} {11}},\ \bibinfo
  {pages} {764} (\bibinfo {year} {2005})}\BibitemShut {NoStop}%
\bibitem [{\citenamefont {Griffiths}\ and\ \citenamefont
  {Walther}(2010)}]{Griffiths2010}%
  \BibitemOpen
  \bibfield  {author} {\bibinfo {author} {\bibfnamefont {A.~J.~V.}\
  \bibnamefont {Griffiths}}\ and\ \bibinfo {author} {\bibfnamefont
  {T.}~\bibnamefont {Walther}},\ }\bibfield  {title} {\textit {\bibinfo {title}
  {{Quantification of carbon contamination under electron beam irradiation in a
  scanning transmission electron microscope and its suppression by plasma
  cleaning}},}\ }\href {\doibase 10.1088/1742-6596/241/1/012017} {\bibfield
  {journal} {\bibinfo  {journal} {Journal of Physics: Conference Series}\
  }\textbf {\bibinfo {volume} {241}},\ \bibinfo {pages} {012017} (\bibinfo
  {year} {2010})}\BibitemShut {NoStop}%
\bibitem [{\citenamefont {Park}\ \emph {et~al.}(2012)\citenamefont {Park},
  \citenamefont {Kwon},\ and\ \citenamefont {Zewail}}]{Park2012}%
  \BibitemOpen
  \bibfield  {author} {\bibinfo {author} {\bibfnamefont {S.~T.}\ \bibnamefont
  {Park}}, \bibinfo {author} {\bibfnamefont {O.-H.}\ \bibnamefont {Kwon}}, \
  and\ \bibinfo {author} {\bibfnamefont {A.~H.}\ \bibnamefont {Zewail}},\
  }\bibfield  {title} {\textit {\bibinfo {title} {{Chirped imaging pulses in
  four-dimensional electron microscopy: femtosecond pulsed hole burning}},}\
  }\href {\doibase 10.1088/1367-2630/14/5/053046} {\bibfield  {journal}
  {\bibinfo  {journal} {New Journal of Physics}\ }\textbf {\bibinfo {volume}
  {14}},\ \bibinfo {pages} {053046} (\bibinfo {year} {2012})}\BibitemShut
  {NoStop}%
\bibitem [{\citenamefont {Cook}\ and\ \citenamefont {Kruit}(2016)}]{Cook2016}%
  \BibitemOpen
  \bibfield  {author} {\bibinfo {author} {\bibfnamefont {B.}~\bibnamefont
  {Cook}}\ and\ \bibinfo {author} {\bibfnamefont {P.}~\bibnamefont {Kruit}},\
  }\bibfield  {title} {\textit {\bibinfo {title} {{Coulomb interactions in
  sharp tip pulsed photo field emitters}},}\ }\href {\doibase
  10.1063/1.4963783} {\bibfield  {journal} {\bibinfo  {journal} {Applied
  Physics Letters}\ }\textbf {\bibinfo {volume} {109}},\ \bibinfo {pages}
  {151901} (\bibinfo {year} {2016})}\BibitemShut {NoStop}%
\bibitem [{\citenamefont {Park}\ \emph {et~al.}(2007)\citenamefont {Park},
  \citenamefont {Baskin}, \citenamefont {Kwon},\ and\ \citenamefont
  {Zewail}}]{Park2007}%
  \BibitemOpen
  \bibfield  {author} {\bibinfo {author} {\bibfnamefont {H.~S.}\ \bibnamefont
  {Park}}, \bibinfo {author} {\bibfnamefont {J.~S.}\ \bibnamefont {Baskin}},
  \bibinfo {author} {\bibfnamefont {O.~H.}\ \bibnamefont {Kwon}}, \ and\
  \bibinfo {author} {\bibfnamefont {A.~H.}\ \bibnamefont {Zewail}},\ }\bibfield
   {title} {\textit {\bibinfo {title} {{Atomic-scale imaging in real and energy
  space developed in ultrafast electron microscopy}},}\ }\href {\doibase
  10.1021/nl071369q} {\bibfield  {journal} {\bibinfo  {journal} {Nano Letters}\
  }\textbf {\bibinfo {volume} {7}},\ \bibinfo {pages} {2545} (\bibinfo {year}
  {2007})}\BibitemShut {NoStop}%
\bibitem [{\citenamefont {Gahlmann}\ \emph {et~al.}(2008)\citenamefont
  {Gahlmann}, \citenamefont {{Tae Park}},\ and\ \citenamefont
  {Zewail}}]{Gahlmann2008}%
  \BibitemOpen
  \bibfield  {author} {\bibinfo {author} {\bibfnamefont {A.}~\bibnamefont
  {Gahlmann}}, \bibinfo {author} {\bibfnamefont {S.}~\bibnamefont {{Tae
  Park}}}, \ and\ \bibinfo {author} {\bibfnamefont {A.~H.}\ \bibnamefont
  {Zewail}},\ }\bibfield  {title} {\textit {\bibinfo {title} {{Ultrashort
  electron pulses for diffraction, crystallography and microscopy: theoretical
  and experimental resolutions}},}\ }\href {\doibase 10.1039/b802136h}
  {\bibfield  {journal} {\bibinfo  {journal} {Physical Chemistry Chemical
  Physics}\ }\textbf {\bibinfo {volume} {10}},\ \bibinfo {pages} {2894}
  (\bibinfo {year} {2008})}\BibitemShut {NoStop}%
\end{thebibliography}%

\end{document}